# Spatiotemporal imaging of nonlinear optics in van der Waals waveguides

Ding Xu[1], Zhi Hao Peng[2], Chiara Trovatello[2], Shan-Wen Cheng[1], Xinyi Xu[2], Aaron Sternbach[3†], Dmitri N. Basov[3], P. James Schuck[2], Milan Delor[1*]

1. Department of Chemistry, Columbia University, New York, NY 10027, United States
2. Department of Mechanical Engineering, Columbia University, New York, NY 10027, United States
3. Department of Physics, Columbia University, New York, NY 10027, United States

† present address: University of Maryland, College Park, MD 20742, United States
*milan.delor@columbia.edu

**Abstract**

Van der Waals (vdW) semiconductors have emerged as promising platforms for efficient nonlinear optical conversion, including harmonic and entangled photon generation. Although major efforts are devoted to integrating vdW materials in nanoscale waveguides for miniaturization, the realization of efficient, phase-matched conversion in these platforms remains challenging. To address this challenge, we develop a far-field ultrafast imaging method to track the propagation of both fundamental and harmonic waves within vdW waveguides with extreme spatiotemporal resolution. Our approach allows systematic optimization of nonlinear conversion by determining the phase-matching angles, mode profiles, and losses in waveguides without *a priori* knowledge of material properties. We focus on light propagation in slab waveguides of rhombohedral-stacked $MoS_2$, an emerging vdW semiconductor with giant nonlinear susceptibility. Our results reveal that these waveguides support birefringent phase-matching, demonstrating the material's potential for efficient on-chip nonlinear optics. This work establishes spatiotemporal imaging of light propagation in waveguides as an incisive and general method to identify new materials and architectures for efficient nonlinear nanophotonics.

**Main**

Nonlinear optical (NLO) processes such as harmonic generation are extensively used in lasers,[1–3] photonic switches[4–8], and quantum information technologies,[9,10] and as probes of material properties including correlated quantum states.[11,12] Van der Waals (vdW) crystals, including transition metal dichalcogenides (TMDs) and ferroelectric materials, have emerged as promising nonlinear optical platforms given their strong light-matter interactions[13,14], large nonlinear susceptibilities[15–18], light-confinement abilities,[19] and versatile fabrication[20,21]. In particular, recent studies on the bulk rhombohedral-phase TMD 3R-$MoS_2$, which comprises nonlinear-dipole-aligned vdW layers with broken inversion symmetry (Fig. 1a), display record nonlinear optical enhancement[22–25]. Indeed, the second-harmonic conversion efficiency of 3R-$MoS_2$ approaches that of lithium niobate over hundred-fold shorter propagation lengths thanks to very large second-order nonlinear susceptibility, $\chi^{(2)}$.[25] The large refractive index dispersion of $MoS_2$, however, leads to low coherence lengths shorter than a micrometer at visible/near-infrared frequencies, beyond which phase-matching considerations become crucial. Recently, a stack-and-twist periodic poling scheme was introduced, achieving quasi-phase-matching over several microns in 3R-$MoS_2$, resulting in 25-fold enhancement of macroscopic efficiency for second harmonic generation (SHG) and spontaneous parametric down-conversion with only few poling periods.[26] The natural next steps are to achieve perfect birefringent phase matching, and to integrate 3$R$-MoS$_2$ in waveguides, allowing simultaneous miniaturization and optimization of NLO conversion efficiencies through precise control over phase- and mode-matching conditions[27–31], including for on-chip applications.

The large refractive index and atomically flat surfaces of TMDs enables facile fabrication of high-quality slab waveguides, and their strong optical anisotropy should allow for birefringent phase matching.[25] These features raise the prospect of using thin slabs of 3$R$-MoS$_2$ as efficient NLO waveguides. Achieving the latter, however, requires precise knowledge of the materials' linear and nonlinear optical properties, loss function and waveguide mode profile, which are difficult to extract with the required precision in realistic geometries to design ideal waveguide structures. Here, we develop a method that empirically accesses these properties by imaging light propagation and SHG within slab waveguides with high spatiotemporal resolution. We show that both multimode and single-mode 3$R$-MoS$_2$ waveguides support phase-matching, paving the way for on-chip integration. Beyond harmonic generation on which we focus, we posit that the unprecedented capability of tracking light propagation and nonlinear conversion in low-loss waveguides with far-field optics will have wide-reaching consequences for our understanding of photon-photon and photon-matter interactions in sub-diffraction structures.

**Direct imaging of fundamental and second harmonic wave propagation in waveguides**

To fabricate slab waveguides, we mechanically exfoliate multilayer flakes of 3$R$-MoS$_2$ and deposit them on a Si/SiO$_2$ substrate (Fig. 1a). Fig. 1b shows an optical image of a 1.25 μm thick slab on which we focus our analysis. The sharp edge identifies the armchair (a.c.) crystal axis, as marked by the white arrows. The as-exfoliated flake forms a multimode Fabry–Pérot cavity. The cavity modes are clearly resolved in the normal-incidence linear reflectance spectrum of the slab (Fig. 1d). Fig. 1c shows the result of exciting the armchair edge of the waveguide with 1030 nm light ($E_{FW}$ = 1.2 eV) at normal incidence and monitoring second-harmonic (SH) light at 515 nm ($E_{SH}$ = 2.4 eV) with conventional far-field microscopy, similar to recent reports.[25,32] The fundamental wave (FW) enters the waveguide by edge scattering. For scattered wavevectors beyond the total internal reflection (TIR) angle of ~20° at the MoS$_2$/SiO$_2$ interface, both the FW and SH waves are waveguided and exit the slab at the opposite edge, as schematically shown in the lower panel of Fig. 1c. Since the waveguide modes are by definition beyond wavevectors accessible by far-field optics, the linear optical image in Fig. 1c does not provide information on light propagation and nonlinear conversion in the interior of the waveguide, obscuring the key NLO processes of interest. We show below, however, that this information can be obtained from nonlinear far-field microscopy.[33,34]

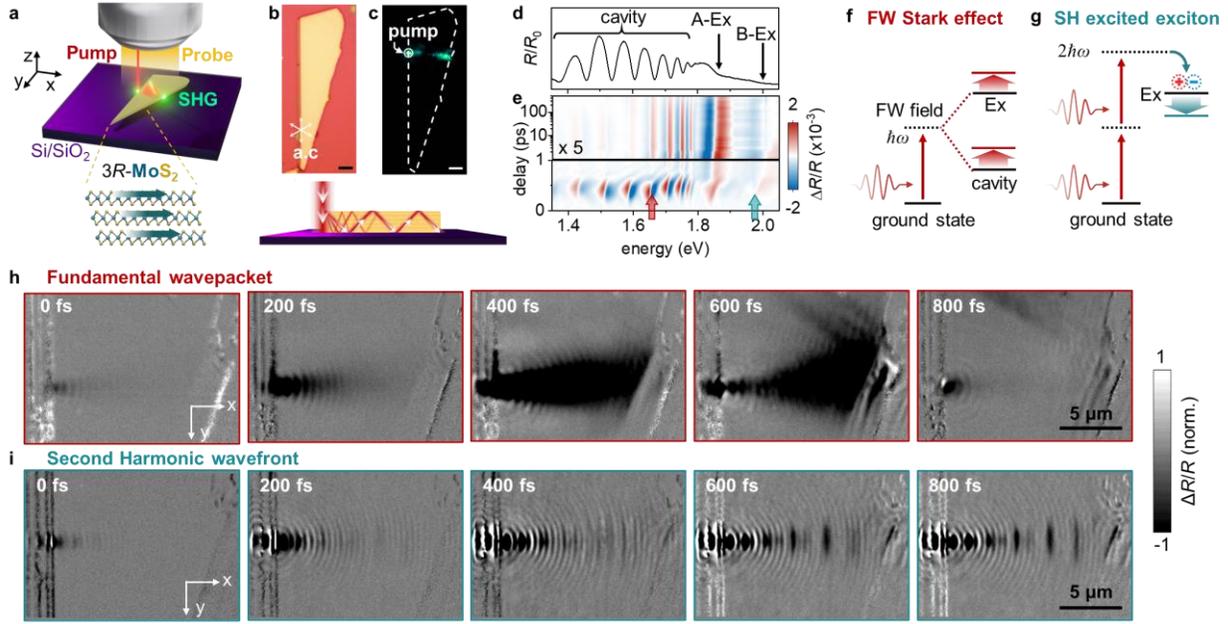

**Figure 1. Imaging the propagation of FW and SH light in 3R-MoS$_2$ waveguides. a**. 3R-MoS$_2$ slab waveguide on Si/SiO$_2$ substrate. The FW is launched by focusing a pump pulse on the left edge. The edge-scattered pump couples perpendicularly into the waveguide and exits at the opposite edge. A time-delayed probe pulse is used to monitor FW and SH propagation in the waveguide. **b.** Optical image of a 1.25 μm thick 3R-MoS$_2$ slab on a Si/SiO$_2$ substrate. The left edge of 3R-MoS$_2$ is cleaved along the armchair (a.c.) direction, marked by the white arrows. **c**. Sample micrograph of the same field of view in (b) showing waveguided SH emission at $E_{SH}$ = 2.4 eV at the entrance and exit edges of the waveguide following edge excitation at $E_{FW}$ = 1.2 eV with a 200 fs pump pulse, as illustrated in the lower inset. **d**. Linear reflectance $R$ of the 3R-MoS$_2$ slab waveguide, normalized to the Si/SiO$_2$ substrate reflectance $R_0$. **e**. Pump-probe transient reflectance spectrum $\Delta R/R$ of the slab waveguide following $E_{FW}$ = 1.2 eV edge excitation, monitored midway between the two waveguide edges with a white light probe pulse. **f**–**g**. Transient signals are dominated by a dynamic stark effect at sub-picosecond time delays (f) and by exciton generation at longer time delays (g). **h**–**i.** Spatiotemporal imaging of FW and SH propagation through the slab waveguide from panel (b), following edge excitation with a 200 fs pump pulse at $E_{FW}$ = 1.2 eV. The incident pump fluence is 5.34 mJ/cm$^2$. The probe energies for FW and SH are at 1.66 eV and 1.96 eV, respectively, indicated by arrows in panel (e). Scale bars are 5 μm.

Instead of directly imaging waveguided FW and SH light at their respective energies, which would require high-momentum coupling such as near-filed optics,[35–39] we instead track the transient changes that these fields impart on the far-field optical properties of the material. We begin by characterizing these changes using pump-probe transient reflectance microscopy (Supplementary Note 1). Fig. 1e shows the transient reflectance of the waveguide upon edge excitation with a diffraction-limited, ~200 fs pump pulse at $E_{FW}$ = 1.2 eV, which is below the optical gap of MoS$_2$ ($E_{opt}$ = 1.85 eV).[40] In the first picosecond following pump excitation, we observe a large transient response, $\Delta R/R$ ~ 0.002, of the Fabry-Pérot cavity modes. All cavity modes blueshift by ~ 0.3 meV due to the presence of the strong FW field in the waveguide, a characteristic signature of the dynamic Stark effect,[41–43] as shown in a detailed analysis in Supplementary Note 2. The amplitude of the Stark signal is proportional to the pump intensity and persists only for a duration close to that of the pump pulse, allowing us to track the FW field by probing the Stark-induced shift in any cavity mode resonance (Fig. 1f).

Once the FW Stark signal subsides (after 1 ps pump-probe delay), a long-lived signal lasting for nanoseconds is observed. The transient spectrum and lifetime of this long-lived signal is almost identical to that obtained upon above-gap photoexcitation of 3$R$-MoS$_2$ (Fig. S5), and is characteristic of photoinduced population of excitons in MoS$_2$.[44,45] The cavity modes also shift because the dielectric function of the material is renormalized[46,47], as shown in Fig. 1e. Crucially, the amplitude of this long-lived signal increases quadratically with pump fluence when exciting below-gap, but linearly with pump fluence when exciting above-gap (Fig. S6) for all probe energies. The quadratic dependence confirms that below-gap excitation cannot directly populate excitons, but does so through a second-order nonlinearity. Indeed, this signal amplitude is directly proportional to the SH light output intensity at the waveguide edge. We therefore assign the long-lived signal to generation of excitons in the waveguide by SH light ($E_{SH}$ = 2.4 eV), which is above gap and is thus able to populate excitons in MoS$_2$ (Fig. 1g). In centrosymmetric 2$H$-MoS$_2$ waveguides of comparable thickness, in which SH generation is suppressed, we indeed observe that the long-lived signal amplitude is 2-3 orders of magnitude smaller for the same pump fluence (Fig. S7). The latter lends strong support to our assignment and excludes possible contributions from two-photon absorption. All together, these measurements indicate that the FW-induced Stark shift and SH-generated exciton signals in stroboSCAT are directly proportional to the FW and SH light intensity (Fig. S6), respectively, serving as distinct far-field reporters of FW and SH light within the waveguide (Fig. 1f-g).

To image the propagation of FW and SH fields within the waveguide, we use far-field stroboscopic scattering microscopy (stroboSCAT), a well-established approach based on transient scattering or reflectance that allows sensitively probing small pump-induced changes to a material's dielectric function.[33,46] A ~200 fs, diffraction-limited pump pulse is launched into the waveguide by edge-scattering, as described above. At controllable time delays, a backscattering monochromatic widefield probe spatially images the pump-induced change in material reflectance (Fig. 1a and Supplementary Note 1). Judicious choice of the probe energies, denoted by arrows in Fig. 1e, allows isolating signal contributions from FW (SH) fields by selecting spectral regions with a strong (weak) Stark response and weak (strong) SH-generated exciton response. Thus, the FW and SH fields can be tracked independently through their distinct effects on the material's far-field optical spectrum.

Key stroboSCAT results are shown in Fig. 1h-i, corresponding to probe energies of 1.66 eV (red arrow in Fig. 1e, tracking FW propagation) and 1.96 eV (green arrow in Fig. 1e, tracking SH propagation), respectively. The striking spatiotemporal evolution is best visualized in Movies S1-S2. Fig. 1h shows how the pump pulse at $E_{FW}$ = 1.2 eV, initially exciting the left edge of the slab, generates a wavepacket that is scattered within the waveguide and propagates perpendicular to the exciting edge at a substantial fraction of light speed (a detailed velocity analysis is provided below). The wavepacket partially bounces back at an angle at the slanted exit edge. The signal is short-lived, reporting only on the spatiotemporal evolution of the FW wavepacket. In contrast, Fig.

1i tracks above-gap SH light through the latter's generation of excitons in MoS$_2$, which are long-lived. Therefore, Fig. 1i shows a wavefront that propagates at the same speed as the FW in Fig. 1h, but leaves a trail of long-lived excitations behind. Our approach thus allows capturing movies of light propagation and nonlinear conversion within waveguides using all-far-field optics.

Both the FW and SH traces in Fig. 1h,i show intensity fringes arising from interference of multiple modes in the multimode waveguide. These fringes are analyzed in detail and reproduced quantitatively using finite element simulations in Supplementary Note 6. The fringe signals are amplified in the SH trace due to the quadratic dependence on FW intensity for SH generation, and provide additional information on phase-matching conditions (*vide infra*). Our approach thus provides a new method to assess the buildup of SH light and depletion of FW within waveguides, in addition to their propagation velocities, which to our knowledge is unprecedented. We show below that such spatiotemporal tracking of waveguide nonlinear optics remarkably allows extraction of phase-matching conditions and waveguide losses without any *a priori* knowledge of material properties.

## Polarization-dependent properties of SH generation in *3R*-MoS$_2$ waveguides

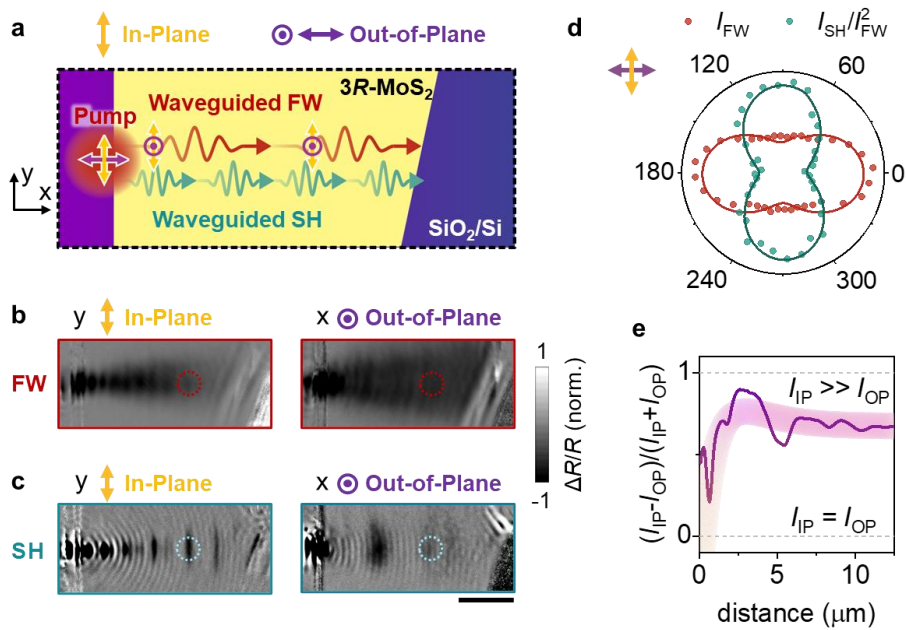

**Figure 2. Anisotropic waveguided FW and SH properties. a**. Pump excitation geometry. Waveguided light have either in-plane (IP, yellow) or out-of-plane (OP, purple) polarizations. **b–c**. Snapshots of waveguided FW (at 400 fs delay) and SH (at 10 ps delay) in IP and OP polarizations, respectively. Scale bar is 5 μm. **d**. The anisotropic FW signal and normalized SHG intensity inside the *3R*-MoS$_2$ slab, measured 10 μm away from the exciting edge, marked by dash circles in **b–c**. **e**. The ratio of SH signal in IP and OP direction as a function of propagation distance. The thick shaded line serves as a visual guide.

To further ascertain the nature of the signals observed in Fig. 1h,i, and understand the anisotropic properties of SH generation in 3R-MoS$_2$ towards achieving birefringent phase-matching, we monitor the FW and SH signals as a function of pump polarization. Previous measurements[25] of light intensity exiting 3R-MoS$_2$ waveguides showed that FW coupling into the waveguide is most efficient when polarized out-of-plane (OP, exciting primarily TM modes in the waveguide). The measurements also showed that SH generation is most efficient when the FW polarization is in-plane (IP, exciting primarily TE modes in the waveguide), i.e. when the electric field of the FW is aligned parallel to the MoS$_2$ slabs. In Fig. 2, we show that stroboSCAT measurements are consistent with this trend, and in addition allow extracting the FW and SH intensity at different locations within the waveguide. The anisotropic properties are evaluated for all pump polarizations at 10º intervals as shown in Fig. S9. Fig. 2b-c show the FW and SH signals, respectively, for IP and OP pump polarizations. The images allow extracting $I_{FW}$ and $I_{SH}$ for each polarization, where $I$ corresponds to the transient reflectance intensity at a given location in the waveguide. Fig. 2d plots the polarization dependence of $I_{FW}$ and of the normalized SH intensity, defined as $I_{SH}/I_{FW}^2 \propto \text{sinc}^2(\frac{\Delta k L}{2})L^2$. The latter accounts for multimode interference in the waveguide that affect local FW intensities. The polar plot confirms that FW waveguide coupling is highest for OP excitation, consistent with Brewster's law[48] and prior measurements[25]. Fig. 2d also shows that the normalized SH intensity is highest for IP polarization of the FW field, concurring with prior linear intensity measurements at the exit edge of the waveguide[25]. These results provide further confidence that the signals tracked with stroboSCAT are valid proxies for the FW and SH light intensities.

In contrast to linear measurements, our approach also allows tracking how the anisotropy evolves as a function of distance from the excitation edge. In Fig. 2e, we plot the anisotropy of the SH intensity defined as $\frac{I_{SH,IP} - I_{SH,OP}}{I_{SH,IP} + I_{SH,OP}}$, where the subscripts IP and OP refer to the polarization of the FW excitation. The anisotropy remains positive and finite throughout the waveguide, indicating that preferential SHG for IP excitation is intrinsic to the system, rather than resulting from the modal structure of the waveguide. The non-monotonic trend in the anisotropy as a function of propagation distance in the waveguide likely results from the interplay of SH generation and SH losses, analyzed below in more detail.

**Birefringent phase matching in bulk waveguides**

We now turn to exploring phase-matching conditions in thick and thin waveguides. Phase-matching requires matching the phase velocities $\omega/k$ of the FW and SH frequencies, where $k = \frac{n\omega}{c}\sin\theta$ is the propagation wavevector, $n$ is the refractive index, $\omega$ is the light angular frequency, $c$ is the speed of light and $\theta$ is the propagation angle. In strongly birefringent media such as MoS$_2$ (see ordinary and extraordinary indices in Fig. S10), the phase velocities can in principle be matched by tuning $\theta$ so that the refractive indices at the FW and SH frequencies are equivalent. Fig. 3a shows the dispersion (energy vs. $k$) for the 1.25 µm waveguide discussed above for TE modes (the TM mode dispersion is shown in Fig. S11b). The experimental dispersion is obtained from angle-resolved reflectance (right side of Fig. 3a) using far-field optics, limited to angles below the numerical aperture of the objective lens (NA = $n\sin\theta$ =1.4). The calculated dispersion (left side of Fig. 3a), obtained from transfer matrix simulations, agrees well with the experimental dispersion, providing confidence in the measured dielectric functions. The simulated dispersion also displays the waveguide modes beyond the TIR line. Under these conditions, the birefringent phase-matching (BPM) angle $\theta^{PM}$ can be approximated using the simple geometric SHG relation (for negative uniaxial crystals):[49]

$$\frac{1}{n_{IP}(\omega)^2} = \frac{\sin^2\theta}{n_{OP}(2\omega)^2} + \frac{\cos^2\theta}{n_{IP}(2\omega)^2} \quad (1)$$

where $n_{IP}(\omega)$, $n_{IP}(2\omega)$ are the ordinary refractive indices of the medium at the FW and SH frequencies, respectively, and $n_{OP}(2\omega)$ is the extraordinary refractive index at the SH frequency. For example, at $E_{FW} = 1.2$ eV, the phase-matching angle is $\theta_{calc}^{PM} = 25.4°$ for 3$R$-MoS$_2$. In Fig. 3b, we plot $\theta_{calc}^{PM}$ (green line) in terms of in-plane wavevector $k_x$ for a range of FW energies. The figure highlights that there is a large range of energies for which BPM occurs within the waveguide regime, i.e. at angles beyond TIR. In thick slabs, multiple waveguide modes (yellow lines) cross this BPM condition as shown in Fig. 3b for our 1.25 μm slab waveguide, suggesting the slab should support waveguide BPM.

For $E_{FW} = 1.2$ eV, we identify a crossing between the 11$^{th}$ order waveguide mode of the 1.25 μm slab and the phase-matching contour in Fig. 3b, highlighted with a red circle. At 1.2 eV, this mode is the lowest-$k$ waveguide mode (beyond the TIR line). To experimentally determine the dominant mode populated by the edge-excitation experiments described above, we leverage the high spatiotemporal resolution of stroboSCAT to analyze the transport velocities of the FW wavepacket and the SH wavefront from Fig. 1h,i (Fig. 3c, with details of the velocity analysis in Fig. S12). By comparing the obtained velocity with the expected group velocity from the dispersion, our time-domain approach provides a reliable measurement of the dominant mode contributing to SHG. The experimental FW velocity for IP excitation at $E_{FW} = 1.2$ eV from Fig. 1h is $v_{IP} = 7.8\%\ c$, as shown in Fig. 3c. The SH wavefront tracks the FW wavepacket at the same speed. An identical analysis for OP excitation provides a velocity of $v_{OP} = 21\%\ c$ (Fig. S11c), consistent with $n_{IP} > n_{OP}$. Fig. 3d shows the expected group velocity extracted from the dispersion, $v_g = \partial\omega/\partial k$, for different modes. For $E_{FW} = 1.2$ eV, only the first waveguide mode matches the experimental velocity, with an expected $v_g = 8\%\ c$. Thus, our spatiotemporal analysis shows that edge excitation primarily populates the lowest-$k$ waveguide mode at 1.2 eV, and that most of the observed SH signal is generated from light propagating within this mode. The momentum of this mode at 1.2 eV is $k_{FW} = 12.6$ μm$^{-1}$, corresponding to $\theta = 24.9°$. This angle matches closely with the above-calculated phase-matching angle of $\theta_{calc}^{PM} = 25.4°$ at 1.2 eV, confirming the possibility of achieving BPM under our experimental conditions.

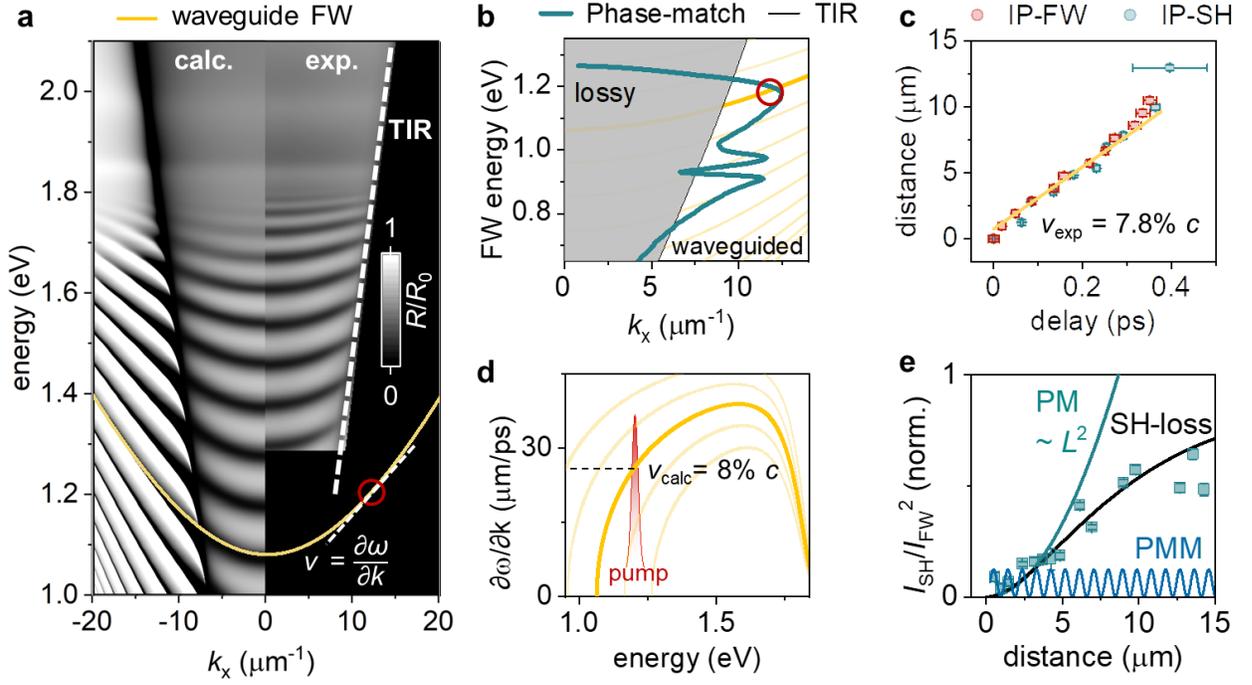

**Figure 3. Determining phase matching conditions through spatiotemporal imaging. a**. Experimental (right) and calculated (left) angle-resolved reflectance spectra showing the TE mode dispersion for a 1.25 μm 3$R$-MoS$_2$ slab. The yellow curve highlights the dispersion of the 11[th] TE mode that dominates the propagation properties of Fig. 1h. The white dashed line marks the total internal reflection line. The full dispersion relations for both polarizations are shown in Supplementary Note 9. **b**. Calculated phase-matching conditions (green), overlaid with TE waveguide modes (yellow) from panel a. The circle highlights the wavevector at which waveguide BPM is observed in our experiments. **c**. FW wavepacket and SH wavefront propagation extracted from stroboSCAT figures shown in Fig. 1h,i. The extracted velocity is 7.8% of light speed. Error bars are one standard deviation. **d.** Group velocities of different modes calculated from the gradient of the mode dispersion. Only one mode at 1.2 eV matches the experimentally-obtained group velocity of ~8% of light speed. At 1.2 eV, this mode corresponds to a momentum $k_x$ = 12.6 μm$^{-1}$, marked by red circle in panel a. **e.** Normalized SH intensity $I_{SH}/I_{FW}^2$ as a function of propagation distance in the waveguide. The black line is a fit using a waveguide SHG model with explicit inclusion of SH losses (see text for details). Lossless phase-matched (PM, $\Delta k = 0$) and phase-*mis*matched (PMM, $\Delta k = 25.2$ μm$^{-1}$, corresponding to normal incidence) cases are indicated with green and blue lines, respectively.

Perfect phase matching implies that the SH intensity increases quadratically with propagation distance $L$ in the nonlinear medium[49]. Since the FW and SH signals in stroboSCAT are directly proportional to the FW and SH light intensities, our measurements allow empirical verification of whether BPM is achieved. In Fig. 3e, we plot the normalized SH intensity as a function of propagation distance, $I_{SH}(L, t)/I_{FW}^2(L, t)$, where the FW and SH intensities are extracted at times $t$ corresponding to $t = L/v$, i.e. tracking the center of the FW wavepacket. The FW signal remains approximately constant throughout the waveguide (Fig. S14), suggesting little depletion, but we observe a clear build-up of normalized SH intensity over the full ~12 μm lateral extent of the waveguide. This intensity buildup extends far beyond the coherence length of 0.94 μm for 3*R*-MoS$_2$ upon normal-incidence excitation (blue line in Fig. 3e),[25] providing strong evidence that at least partial phase-matching is achieved in our structures. Nevertheless, beyond ~7 μm, the normalized SH intensity exhibits a deviation from the $L^2$ proportionality expected for perfect BPM. This deviation is expected since the SH light at 2.4 eV is above the optical gap of 3*R*-MoS$_2$ and is therefore partially absorbed by the material.[50] We employ a system of coupled first-order differential equations describing SHG in lossy waveguides to model our results (simulation details are provided in Supplementary Note 13).[51] The black line in Fig. 3e shows the result of this model, which uses two fitting parameters (the SHG efficiency and loss rate) to fit three curves globally. We reach close agreement with the experimental results when assuming an SH absorption coefficient of $\tau^{-1}_{\text{waveguide}} = 0.7$ μm$^{-1}$ for 3*R*-MoS$_2$. This value is much smaller than the absorption coefficient of MoS$_2$ at normal incidence at 2.4 eV, $\tau^{-1} \approx 10^2$ μm$^{-1}$.[52] This weak absorption of SH light can be at least partially rationalized by considering that in a BPM scheme starting with ordinary FW polarization, the SH polarization is extraordinary; since the primary optical transition dipole in MoS$_2$ is polarized in-plane, SH light is only weakly absorbed in the waveguide.[53] Overall, these results provide strong evidence that we achieve BPM in slab waveguides of 3*R*-MoS$_2$, even with simple coupling schemes such as edge excitation.

**Polariton-assisted modal phase matching in thin waveguides**

Although the thick waveguide described above provides a convenient platform to achieve phase-matching due to a large density of waveguide modes, the benefits of waveguides are better realized in thin waveguides, where strong light confinement enables large nonlinear enhancements.[54–58] To investigate this regime, we now turn to a single-mode 3*R*-MoS$_2$ slab waveguide of 154 nm thickness (Fig. 4a). Fig. 4b shows the SH emission collected in the far field following $E_{\text{FW}} = 1.2$ eV edge excitation, confirming that waveguiding can be achieved in these thin flakes similarly to bulk flakes. Fig. 4c overlays the calculated dispersion of the TE mode at the FW frequency (traced with a red dashed line), with the TM modes at the SH frequency (traced with green dashed lines). Modal phase-matching is achieved in these structures at frequencies and momenta where the TE$_{\text{FW}}$ (red) and TM$_{\text{SH}}$ (green) lines cross in Fig. 4c, corresponding to a matching of effective refractive indices. Although other modal phase-matching conditions exist, we focus on this specific combination because it maintains TM orientation of the SH light to minimize losses. We identify two modal phase-matching conditions at $E_{\text{FW}} = 0.96$ eV and 0.90 eV. Importantly, these modal crossings are facilitated by a flattening of the TM dispersion as the SH energy approaches the A- and B-exciton resonances of MoS$_2$. This dispersion flattening is caused by strong interactions between photons and excitons, resulting in a renormalization of light and matter eigenstates into hybrid quasiparticles known as exciton-polaritons. Such hybridization is readily achieved in MoS$_2$ slabs.[59] A characteristic signature of polaritons is an anti-crossing in the dispersion between the photon and exciton modes. When two exciton bands are present, the anti-crossing results in lower, middle and upper polariton branches.[60–62] The two modal phase-matching conditions in Fig. 4c correspond to the TE$_{\text{FW}}$ branch crossing the lower and middle polariton branches formed by hybridization of the TM$_{\text{SH}}$ with the A- and B-excitons of MoS$_2$, respectively.

Such polariton-assisted modal phase-matching in waveguide geometries remains, to our knowledge, an underexplored field.

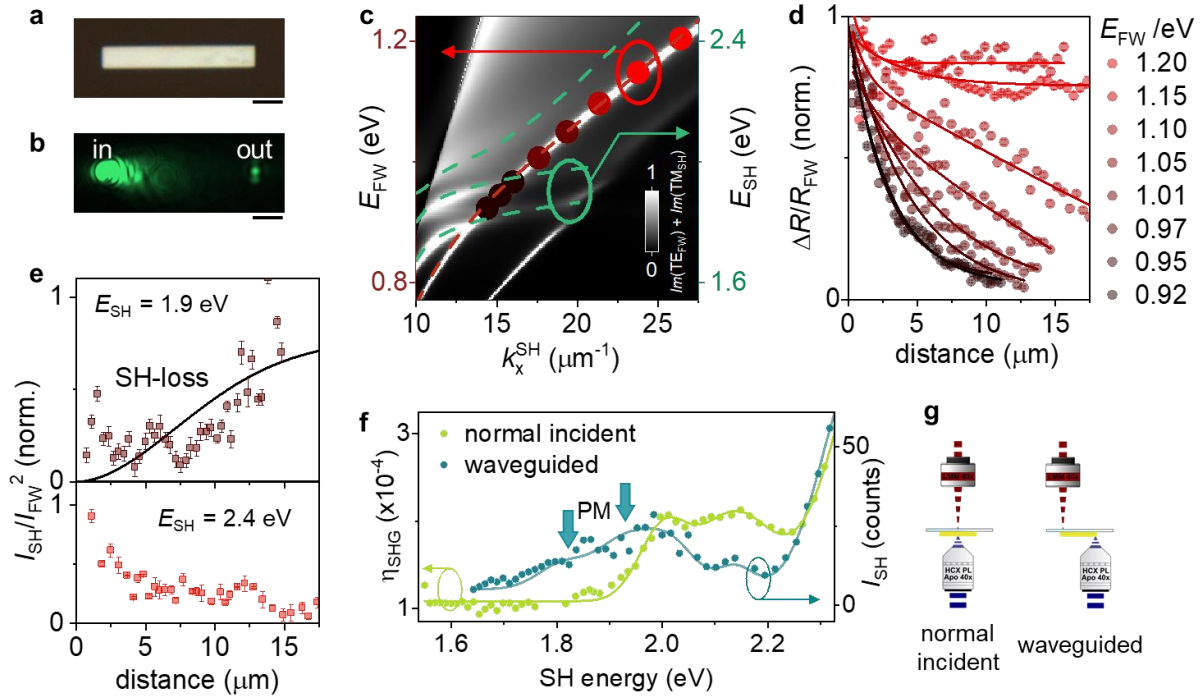

**Figure 4. Modal phase matching in thin waveguides. a.** Optical image of a 154 nm thick single-mode 3$R$-MoS$_2$ waveguide on a glass substrate. **b**. Waveguided SH emission at $E_{SH}$ = 2.4 eV collected by a far-field objective. Scale bars are 5 μm. **c**. Dispersion of FW (red, TE mode) and SH (green, TM mode) in the 154 nm thick single-mode waveguide, extracted from transfer matrix simulations. This figure overlays the FW modes onto the SH modes by rescaling the energy and momentum axis so that mode crossings correspond to matched phase velocities. **d**. FW signal as a function of $L$ in the waveguide extracted from stroboSCAT (raw data in Fig. S17), with FW energies spanning $E_{FW}$ = 0.92-1.20 eV. Substantial FW depletion is observed near $E_{FW}$ = 0.96 eV due to efficient phase-matched SH conversion. **e**. SH buildup extracted from stroboSCAT (raw data in Fig. S17b) for phase-*mis*matched and phase-matched SH generation at $E_{SH}$ = 2.4 eV and 1.9 eV, respectively. The black line is a fit using a waveguide SHG model with explicit inclusion of SH losses. Error bars are one standard deviation. **f**. SH emission intensity measured under waveguide (green) and normal-incidence (light green) geometries in the same slab. The pump power is 2 mW for all FW energies, using ~60 fs pulses. The arrows indicate the two modal phase-matching energies. **g**. Geometries used for the data shown in panel f.

Using stroboSCAT, we first evaluate the evolution of the FW signal propagating in the thin waveguide, for the different excitation energies highlighted with filled circles in Fig. 4c. Fig. 4d displays the extracted FW signal as a function of propagation. At frequencies that are phase-*mis*matched (e.g. $E_{FW}$ = 1.2 eV), the FW signal is approximately constant throughout the waveguide (Fig. S17a), indicating low losses and negligible FW intensity depletion, in line with our observations in multimode waveguides described above. However, as $E_{FW}$ approaches 0.92 eV, the FW signal exhibits substantial depletion (Fig. 4d, raw data is shown in Fig. S17b) despite being far below the optical gap of $MoS_2$. We observe almost complete depletion of the FW signal over a propagation distance of 10 μm at $E_{FW}$ = 0.95 eV. This depletion suggests either highly efficient and phase-matched nonlinear conversion of the FW into SH light, or strong losses of the FW. We emphasize that we do not observe this substantial FW depletion in the thick waveguide described in Fig. 3, even for identical pump energies (Fig. S18), confirming that there should be no direct absorption losses from $MoS_2$ at these below-gap FW energies. These results indicate that strong confinement and exciton- or polariton-mediated enhancement of nonlinear interactions[63–65] are likely responsible for the observed FW signal depletion. Directly tracking FW depletion in waveguides through spatiotemporal imaging thus provides an effective way to characterize nonlinearities and identify phase-matching conditions. Furthermore, frequency-domain analysis of periodic fringes observed in stroboSCAT signals in single-mode waveguides under phase-mismatched conditions provides a direct measurement of the degree of phase-mismatch (Supplementary Note 16), an additional benefit of our approach.

Fig. 4e plots the normalized SH signal intensity $I_{SH}/I_{FW}^2$ extracted from stroboSCAT data for $E_{FW}$ = 1.2 eV (phase mismatched) and 0.95 eV (phase matched). We observe no SH build-up in the phase-*mis*matched case, as expected given the short coherence length. In the phase-matched case, however, we observe an increase of the normalized SH intensity over 15 μm, though we emphasize that the absolute SH signal begins to drop after 3 μm due to FW depletion (Fig. S17b). We model this behavior using the same system of coupled equations as that used in Fig. 3e, with the fit result shown as a black line in Fig. 4e (details in Supplementary Note 13). Although the data exhibit features not captured in the model, the fit suggests that strong depletion is caused by a combination of efficient nonlinear conversion and SH absorption. The best fit is obtained when using a 7-fold larger $|\chi^{(2)}|^2$ compared to the multimode system of Fig. 3e, and a slightly lower SH loss of $\tau_{waveguide}^{-1}$ = 0.48 μm$^{-1}$. The larger $\chi^{(2)}$ compared to thick waveguides is likely mediated by confinement and polariton-assisted enhancement, whereas the lower loss likely reflects the fact that a substantial fraction of the light field extends beyond the thin waveguide.

To further ascertain that these results are consistent with more traditional measurements of SH generation, we compare the SH light intensity directly collected in the far-field for a variety of pump wavelengths in either waveguide or transmission geometries in the same slab (Fig. 4f). The experimental geometries are illustrated in Fig. 4g, using reflective optics for FW excitation to access a broad frequency range. Given the observed FW depletion and saturation of the SH signal at around $L$ = 3 μm in stroboSCAT measurements, we laser-cut the waveguide into a 3 μm × 4 μm rectangle, allowing us to measure the SH light output at the waveguide edge near its maximum intensity. In the transmission geometry, the SH efficiency $\eta_{SHG}$ is quantified by measuring the transmitted SH emission power $P_{SH}$ normalized to the input pump laser power $P_{FW}$ ($\eta_{SHG}$ = $P_{SH}/P_{FW}$). In the waveguide geometry, only a small fraction of the FW is coupled into the waveguide, and a small fraction of the emitted SH light is collected in the far field, limiting our comparison between the two geometries to relative efficiencies. The normal-incidence transmission measurements (in which no phase-matching is possible) in Fig. 4f are consistent with prior measurements in ultra-thin 3R-$MoS_2$ flakes,[25] with $\eta_{SHG}$ showing peaks at the B-exciton (not at the A-exciton) and slightly above the B-exciton, and then rising rapidly at higher energies due to nesting of electronic bands in MoS$_2$ that is known to drastically enhance nonlinearities.[66,67] In the waveguide geometry, there are clear distinctions in the low-energy region. Most strikingly, we observe two peaks at energies near the predicted modal phase-matching conditions of Fig. 4c ($E_{SH}$ = 1.9 eV and 1.8 eV, highlighted with arrows in Fig. 4f). There is no measurable SHG in the normal-incidence data at these energies, indicating a drastic SHG enhancement in this energy range in the waveguide geometry. These observations corroborate our stroboSCAT measurements, indicating the possibility of achieving phase matching even in highly confined slab waveguides of 3$R$-MoS$_2$.

**Conclusion**

We have developed a new approach based on nonlinear far-field microscopy to image light propagation and nonlinear optical conversion in low-loss waveguides with femtosecond and sub-micrometer spatiotemporal resolution. We demonstrate the power of our approach on slab waveguides of 3$R$-MoS$_2$, a material with large nonlinear susceptibility in the vanguard of current interest. We show that we can predict and realize phase-matching for enhanced nonlinear conversion in both multimode and single-mode slab waveguides. Our approach allows direct extraction of phase-matching angles or degree of phase-mismatch, waveguide mode profile, absorption and waveguide losses, and relative nonlinear conversion efficiencies without *a priori* knowledge of material properties. Ongoing efforts to achieve better in- and out-coupling of light in these waveguides, for example through grating couplers or tapered fibers,[32,68] will firmly establish 3$R$-MoS$_2$ as ideal components for on-chip nonlinear optics. More generally, this work offers a powerful new imaging approach for understanding the fundamental properties of light in complex photonic architectures and processes, including wave evolution dynamics[69–71], high-harmonic generation[11,12], and time reversal transformations[72,73].

**Method**

The details of the optical measurements, analysis and model fits are provided in the Supplementary Information.

Spatiotemporal transport of the waveguided FW and SH within slab waveguides are measured using ultrafast stroboscopic scattering microscopy and transient reflectance microscopy. The optical systems are illustrated schematically in Fig. S1. A Yb:KGW ultrafast regenerative amplifier (Light Conversion Carbide, 40 W, 1030 nm fundamental, 1 MHz repetition rate, ~200 fs pulsewidth) seeds an optical parametric amplifier (OPA, Light Conversion, Orpheus-F) to generate an idler pulse tunable from 1.045 to 2.907 um with ~50 fs pulse width. The 1030 nm fundamental or idler serve as pump pulses, and a super-continuum white light generated on a YAG window (EKSMA, 555-712, ø 12.7 mm, 5 mm) is used as probe. The pump and probe beams are directed to a home-built microscopes equipped with a high numerical-aperture oil-immersion objective (Leica HC Plan Apo 63x, 1.4 NA oil immersion). The pump beam focuses on the sample plane to a diffraction-limited spot. The probe beam focuses on the back focal plane of the objective by an $f$ = 250 mm widefield lens, enabling widefield illumination of the sample. The reflected and scattered fields are imaged on a CMOS camera (Blackfly S USB3, BFS-U3-28S5M-C). Further details of the ultrafast stroboSCAT system are described in our previous reports[46,74].

The linear transmittance and reflectance measurements are obtained using a different optical configuration shown in Fig. S2. A reflective objective is used for the excitation to enable broader spectral coverage. The pump laser is focused by a reflective objective (Thorlabs LMM40X-P01, Infinity Corrected 40x, 0.5 NA). The emitted SHG is detected by a transmissive air objective (Leica HCX PL Apo 40x, 0.85 NA, air). The light source is the same as that described above. The SH emission is detected using a cooled CMOS camera (Thorlabs CC505MU). The power of the SHG emission in the transmitted geometry is measured using a power meter (Thorlabs PM100D) to extract the absolute conversion efficiency.

**Acknowledgements**
This work was primarily supported by Programmable Quantum Materials, an Energy Frontier Research Center funded by the US Department of Energy, Office of Science, Basic Energy Sciences, under Award DE-SC0019443. Instrument development was supported by the National Science Foundation under Grant Number CHE-2203844 (M.D.). Linear optical characterization was supported by the Arnold and Mabel Beckman Foundation through a Beckman Young Investigator award. D.X. acknowledges a Kathy Chen Fellowship. C.T. acknowledges the European Union's Horizon Europe research and innovation programme under the Marie Skłodowska-Curie PIONEER HORIZON-MSCA-2021-PF-GF grant agreement No 101066108.

**Author contributions**
D.X., P.J.S., and M.D. conceived and designed the experiment. D.X. built the optical instruments and acquired the experimental data. Z.P., X.X., C.T. and S.C. prepared the samples. Z.P. and D.X. performed the linear microscope measurement of the second-harmonic generation. D.X. and A.S. analyzed the transport data and implemented the numerical models. M.D., P.J.S. and D.N.B. supervised the study. D.X. and M.D. wrote the article with input from all authors.

**Competing interests**
The authors declare no competing interests.

# Supplementary Information for:

# Spatiotemporal imaging of nonlinear optics in van der Waals waveguides

Ding Xu[1], Zhi Hao Peng[2], Chiara Trovatello[2], Shan-Wen Cheng[1], Xinyi Xu[2], Aaron Sternbach[3†], Dmitri N. Basov[3], P. James Schuck[2], Milan Delor[1*]

1. Department of Chemistry, Columbia University, New York, NY 10027, United States
2. Department of Mechanical Engineering, Columbia University, New York, NY 10027, United States
3. Department of Physics, Columbia University, New York, NY 10027, United States

[†] present address: University of Maryland, College Park, MD 20742, United States

*milan.delor@columbia.edu

Table of contents

# 1. Far-field ultrafast optical microscopy

The spatiotemporal transport of the waveguided fundamental wave (FW) and second-harmonic wave (SH) within $3R$-MoS$_2$ are measured using ultrafast stroboscopic scattering microscopy (stroboSCAT) and transient reflectance microscopy. The systems used here are described in detail in prior publications[1,2] and illustrated schematically in Figure S1.

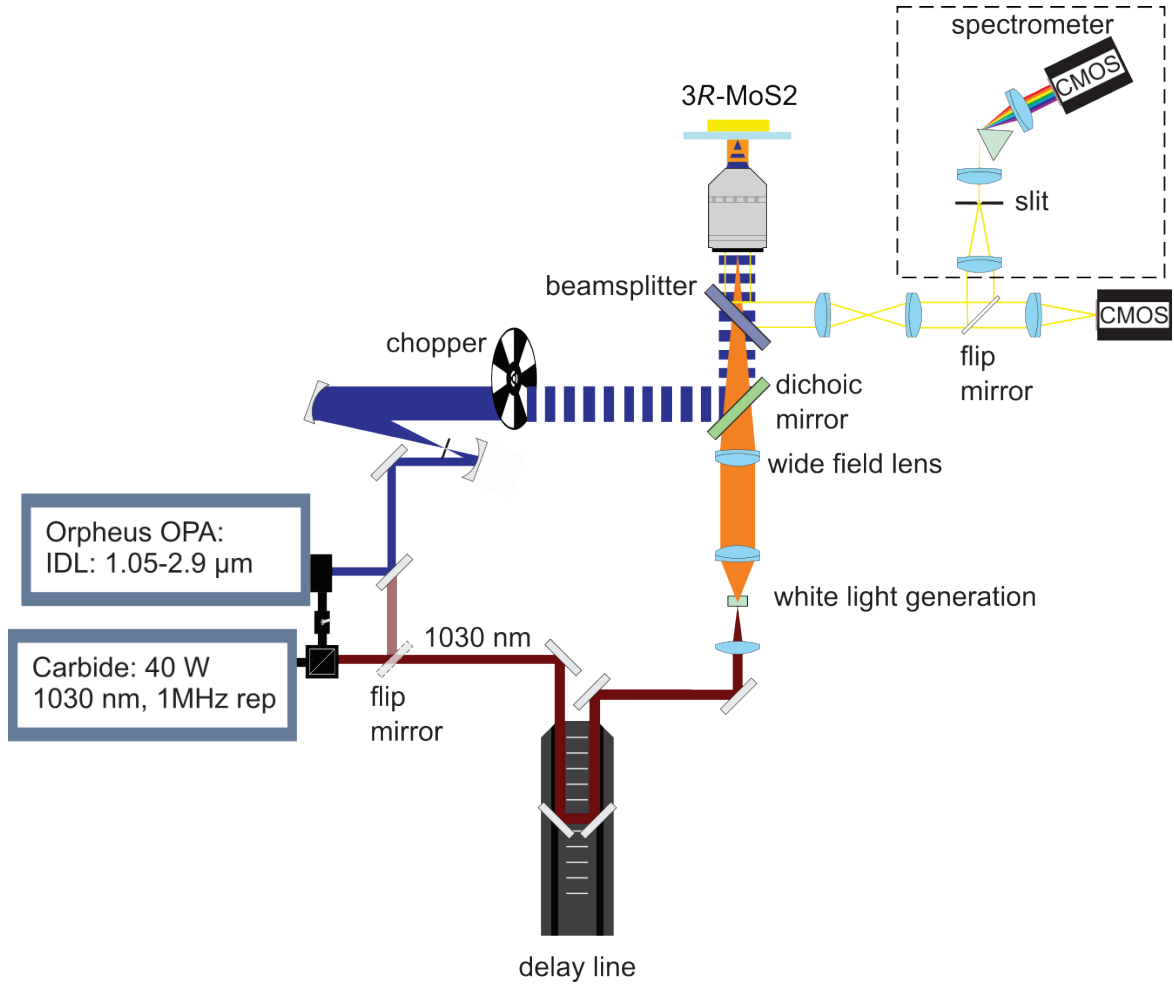

Figure S1. stroboSCAT setup schematic. A Yb:KGW ultrafast regenerative amplifier (Light Conversion Carbide, 40 W, 1030 nm fundamental, 1 MHz repetition rate) seeds an optical parametric amplifier (OPA, Light Conversion, Orpheus-F) with a idler tunable from 1.045 to 2.907 um. The 1030 nm fundamental or idler serve as pump pulses, and a super-continuum white light generated on YAG window (EKSMA, 555-712, ø 12.7 mm, 5 mm) is used as probe. The pump and probe beams are directed to the high numerical-aperture oil-immersion objective (Leica HC Plan Apo 63x, 1.4 NA oil immersion). The pump beam is collimated before the objective lens and focuses on the sample plane. The probe beam focuses on the back focal plane of the objective by an $f = 250$ mm widefield lens, enabling widefield illumination of the sample. The reflected probe light at the sample interface with the back-scattered light from the sample forms images on a CMOS camera (Blackfly S USB3, BFS-U3-28S5M-C). Further details of the ultrafast stroboSCAT system are described in our previous reports[1,2].

The data for Figure 4f as well as the linear transmittance and reflectance measurements are obtained using a different optical configuration shown in Figure S2. A reflective objective is used for the excitation to enable broader spectral coverage.

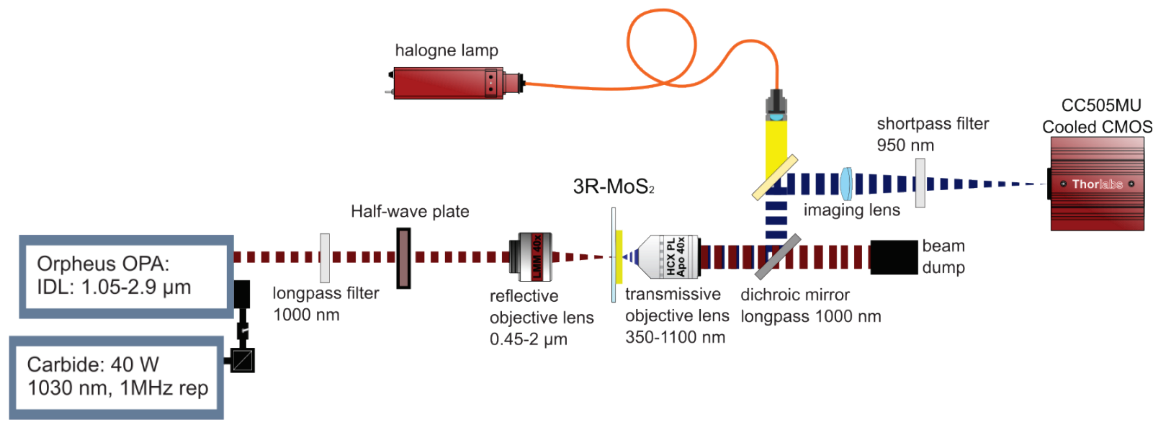

Figure S2. Second-harmonic emission microscope schematic. The pump laser is focused by a reflective objective (Thorlabs LMM40X-P01, Infinity Corrected 40x, 0.5 NA). The emitted SHG is collected by a transmissive air objective (Leica HCX PL Apo 40x, 0.85 NA, air). The light source is the same as that described in Figure S1. The SHG emission is detected using a cooled CMOS camera (Thorlabs CC505MU). The absolute power of the SHG emission in the transmitted geometry is measured using a power meter to extract the absolute conversion efficiency.

## 2. Analysis of the dynamic Stark effect used to image FW propagation

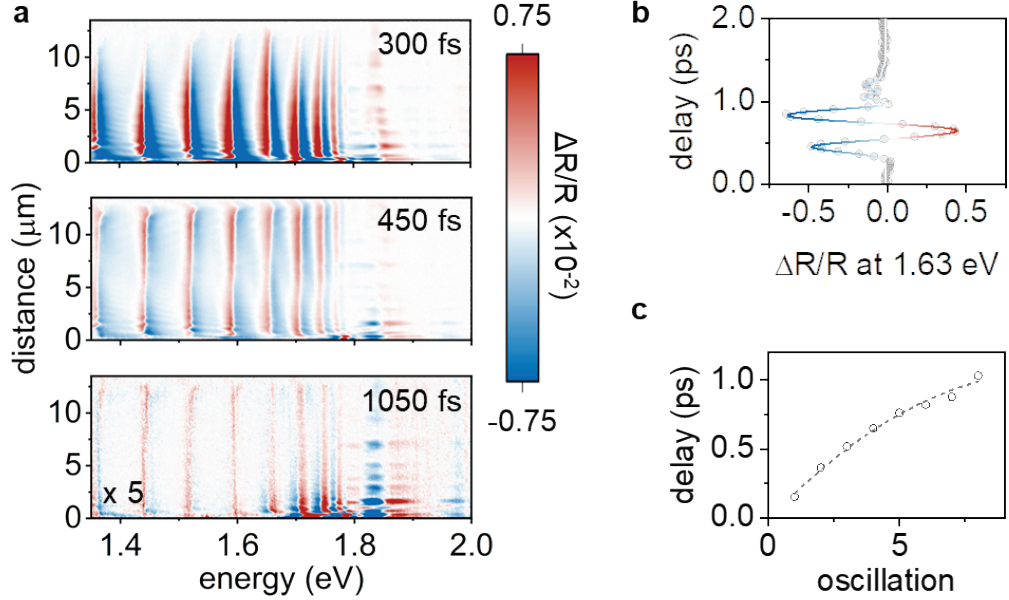

Figure S3. (a) Spatiotemporally resolved transient reflectance spectrum. In each panel, the *x*-axis and the *y*-axis represent the spectral dimension and spatial dimension, respectively. At every delay time, the waveguided FW propagates from the bottom of the panel to the top. (b–c) Oscillations associated with the Stark effect at a cavity resonance at $E = 1.63$ eV. Panel b shows that the transient reflectance $\Delta R/R$ of FW fluctuates between negative and positive values at least three times before 2 picoseconds. Panel c shows the delay times corresponding to the peaks of each oscillation.

We can model the dynamic Stark effect in excitonic materials in cavities using the photon-exciton coupling Hamiltonian[3,4]:

$$H = \begin{bmatrix} U_{ex} & \hbar\Omega_{cav} & \hbar\Omega_{FW} \\ \hbar\Omega_{cav} & \hbar\omega_{cav} & 0 \\ \hbar\Omega_{FW} & 0 & \hbar\omega_{FW} \end{bmatrix} \quad (1)$$

where $U_{ex}$ is the energy of exciton states, $\omega_{cav}$ and $\omega_{FW}$ are the angular frequencies of cavity mode photons and FW photons, respectively, and $\Omega_{cav}$ and $\Omega_{FW}$ correspond to their Rabi frequencies. The Rabi splitting of exciton-cavity photon coupling is measured as $\hbar\Omega_{cav} = 142$ meV using a coupled oscillator model; the FW field Rabi splitting is measured as $\hbar\Omega_{FW} = 0.3$ meV by fitting the transient blueshift of cavity spectral lines in transient reflectance (TR) spectra in figure S4. Diagonalization of the Hamiltonian (Equation 1) indicates that the transient spectrum shift depends on the energy of the FW field with respect to the exciton state. Under our experimental conditions, the FW field at $E_{FW} = 1.20$ eV is lower in energy than the exciton at $U_{ex} = 1.85$ eV, resulting in an expected Stark-induced blueshift of the signal, as observed. Another feature of the dynamic Stark effect is the coherent perturbation of pump-probe polarization at early delay time,[5] manifesting in the alternating phase of the TR signal between 0.5-1 picosecond in Figure S3. This effect was also observed in quantum well exciton-polariton systems.[4]

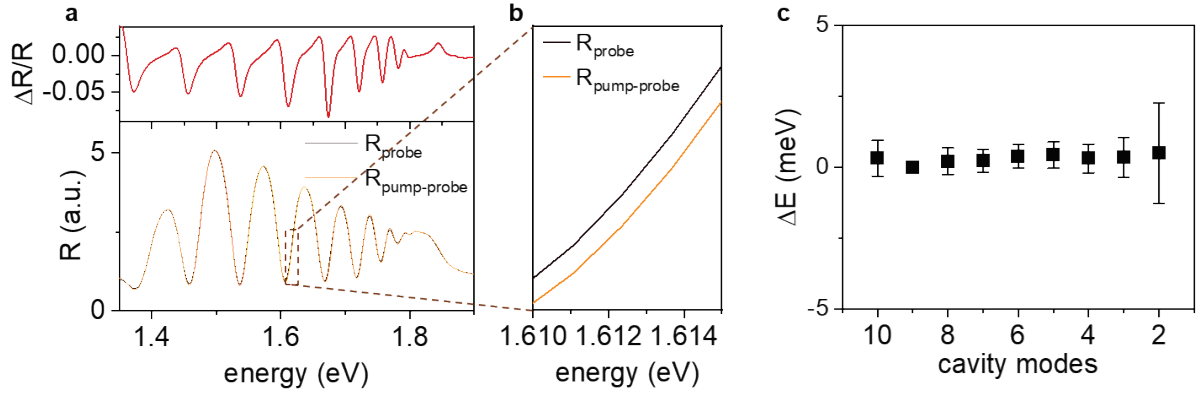

Figure S4. (a) TR spectrum of Stark effect at 0.5 ps delay. The upper panel shows the TR contrast $\frac{\Delta R}{R} = \frac{R_{\text{pump-on}} - R_{\text{pump-off}}}{R_{\text{pump-off}}}$. The lower panel shows the absolute reflectance spectra $R_{\text{pump-off}}$ obtained from linear reflectance and $R_{\text{pump-on}} = \frac{\Delta R}{R} * R_{\text{pump-off}} + R_{\text{pump-off}}$ in yellow and black lines, respectively. (b) Enlarged figure of the plot in panel a. (c) Experimental blueshift of Stark effect for different cavity modes from 1.4 eV to 1.75 eV. Error bars are one standard deviation.

## 3. Second-harmonic wave excitation of excitons and associated transient reflectance features

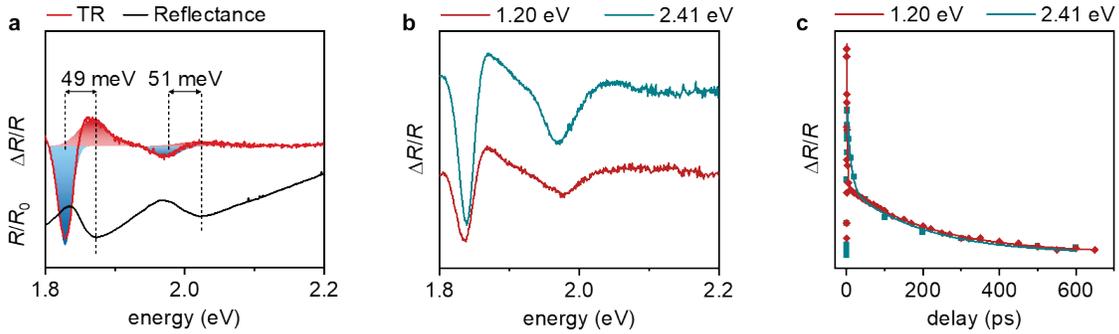

Figure S5. (a) Redshift of the A- and B-exciton transitions observed in in transient reflectance spectra (here plotted at 5 ps time delay) following below-gap pump excitation at 1.20 eV. The linear reflectance of 3$R$-MoS$_2$ is shown below for reference (black line). The latter shows the characteristic absorption (dip in reflectance) peaks corresponding to A- and B-excitons at $E_{A\text{-}Ex}$ = 1.88 eV and $E_{B\text{-}Ex}$ = 2.03 eV. The TR spectrum shows ground-state bleaching at A- and B-exciton resonances (peaks highlighted in red) and photon-induced absorption at lower energies (dips highlighted in blue). The photoinduced absorption redshift is characteristic of bandgap renormalization[6], resulting here in an apparent redshift of ~50 meV. These spectra indicate that excitons or free carriers are populated by below-gap excitation, assigned to SH generation based on the fluence dependence reported in Figure S6. (b) Transient reflectance spectra upon pump excitation at 1.20 eV (red, below-gap) and 2.41 eV (green, above-gap) at 500 ps delay time. The spectra are offset for clarity. The similarity between the two spectra supports the hypothesis that below-gap excitation results in SH light absorbed by the material, effectively resulting in 2.41 eV excitation. (c) The bandgap renormalization signals for 1.20 eV (red) and 2.40 eV (green) excitation have characteristic lifetimes of 224 ps and 206 ps, respectively.

## 4. Linear and quadratic fluence dependence of FW and SH signals

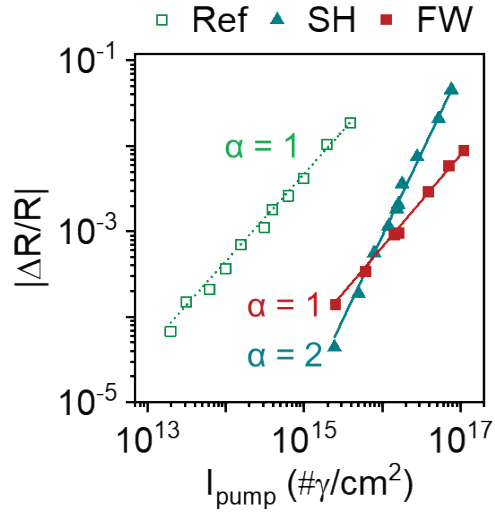

Figure S6. Transient reflectance $\Delta R/R$ of FW and SH as a function of incident pump power (plotted as number of photons/cm$^2$) measured at delay times of 500 fs (FW signal) and 100 ps (SH signal). In this experiment, the pump energy is at $E = 1.2$ eV (below gap), and the probe energy is near the A-exciton resonance at $E = 1.82$ eV. The FW (red solid line) signal increases linearly with pump power. The SH (green solid line) increases quadratically with pump power. The first- and second-order power dependencies of the FW and SH signals confirm the different nature of interactions, as discussed in the main text. The FW signal arises from the Stark effect, whereas the SH signal arises from SH-generated excitons. To confirm the accuracy of the pump-power dependent nonlinearity measurements, we conducted a parallel experiment with pump energy set to match the SH energy at $E_{SH}=2.4$ eV, corresponding to above-bandgap pumping (green dashed line). This above-gap pump directly excites excitons or free carriers in the material. The transient reflectance signal in this case grows linearly with pump power, as expected. We note that the quadratic power dependence observed for the green solid line could correspond to either SH-generated excitons or two two-photon absorption. We rule out that two-photon absorption dominates the signal in Figure S7.

## 5. Fluence dependence of transient reflectance signal in 2H and 3R-MoS₂

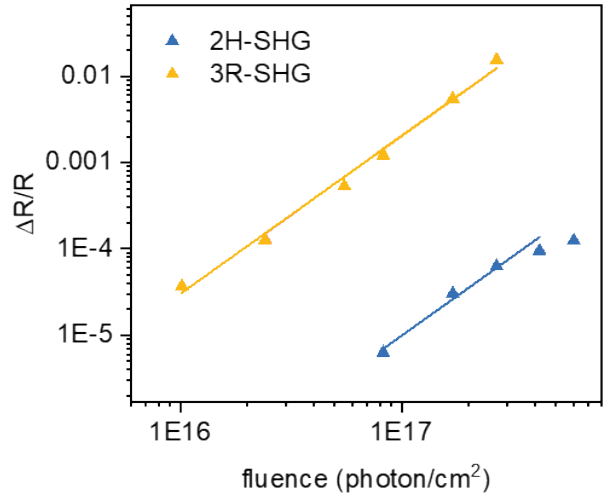

Figure S7. Transient reflectance Δ$R/R$ as a function of pump power at a delay time of 100 ps in 2$H$-MoS$_2$ (blue, thickness is 1.51 μm) and 3$R$-MoS$_2$ (yellow, thickness is 1.25 μm) slab waveguides. Pump energy is at $E = 1.203$ eV, and the A-exciton resonant probe energy is at $E = 1.85$ eV. In 2$H$-MoS$_2$, SHG is drastically suppressed due to symmetry, but the susceptibility tensor elements associated with two-photon absorption remain essentially unaffected.[7,8] The comparison between signals in 2$H$ and 3$R$ slabs thus allows us to distinguish contributions from SHG and two-photon absorption. Our results show that Δ$R/R$ increases quadratically with pump power for both samples. However, the signal in 3$R$-MoS$_2$ is 2-3 orders of magnitude higher than in 2$H$-MoS$_2$. The signal in 2$H$-MoS$_2$ could be due to residual two-photon absorption, setting an upper limit for the contribution of two-photon absorption in the signal in 3$R$-MoS$_2$ to ~1%. We also observe a saturation effect in 2$H$-MoS$_2$ that isn't present in 3$R$-MoS$_2$. These results provide strong evidence that the signal assigned to SH-generated excitons in the text is indeed strongly dominated by SHG rather than two-photon absorption.

## 6. COMSOL simulation of multimode waveguide interference

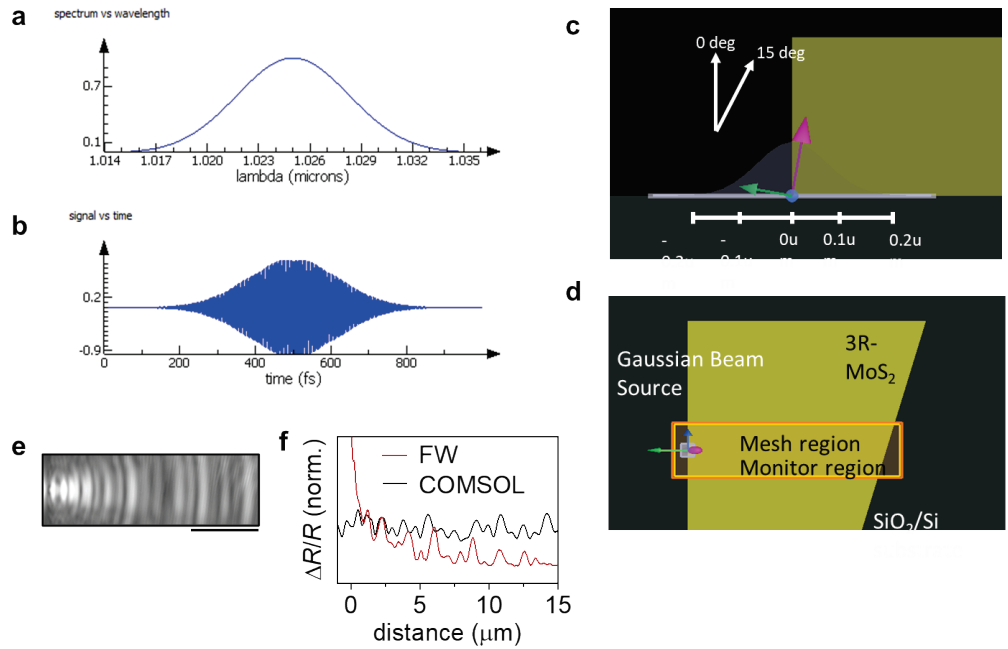

Figure S8. COMSOL simulation of FW field for a 3R-MoS$_2$ waveguide slab with a thickness of 1.25 μm. (a–b) Fundamental mode electric field initialization conditions. The fundamental mode is a laser pulse at 1.025 μm with spectral width of 8 nm. The pulse duration is 200 fs. (c–d) Transport initialization conditions of the fundamental mode in the 3R-MoS$_2$ waveguide slab. The fundamental mode pulse propagation starts at the exciting edge with the transverse polarization direction at an angle of 15 degree to the edge in the vertical direction. The electric field distribution in the 3R-MoS$_2$ waveguide slab is calculated within a 5×20 μm mesh region. (e) Simulation result of electric field profiles in the horizontal directions. (f) The simulation results for the fundamental mode reveal fringe signals (black line) that are consistent with those observed in the experimental spatial transport profile of the fundamental mode, as measured by stroboSCAT (red line). These fringes are due to mode interference in the multimode waveguide.

## 7. Polarization dependent FW and SH signals.

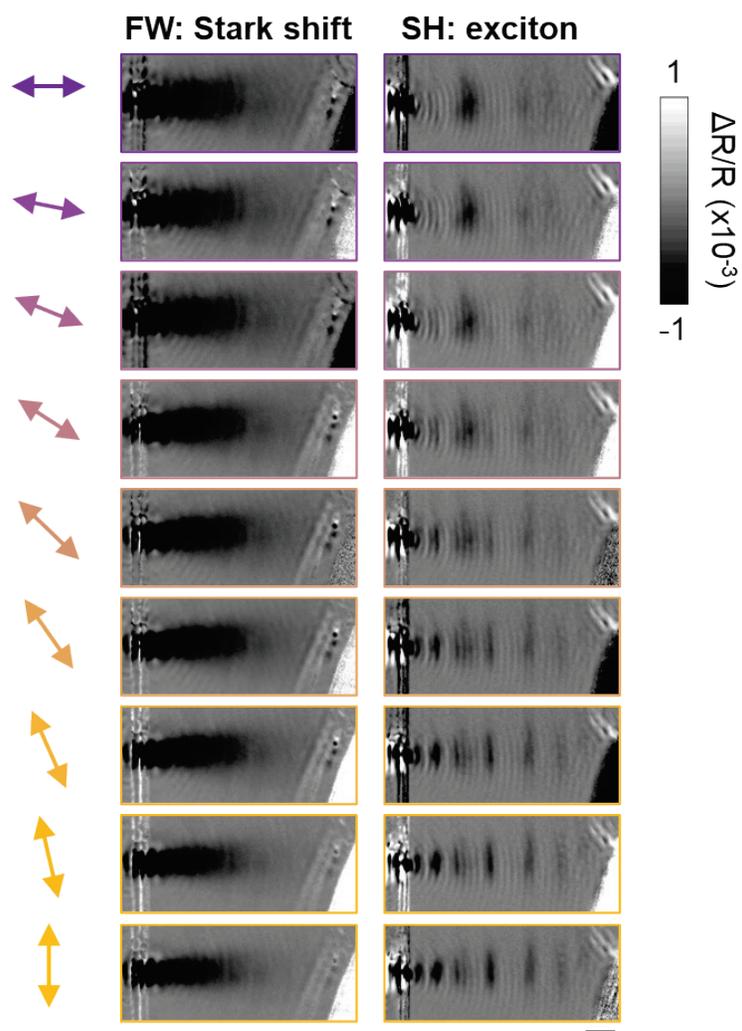

Figure S9. Raw data for Figure 2. Polarization-dependent FW and SH signals. FW and SH are probed at 1.66 eV and 1.96 eV, respectively, and plotted for delay times of 400 fs and 10 ps, respectively. Pump fluence is 5.34 mJ/cm$^2$. From top to bottom, the polarization of FW changes from OP (purple) to IP (yellow).

## 8. Anisotropic dielectric function of 3$R$-MoS$_2$

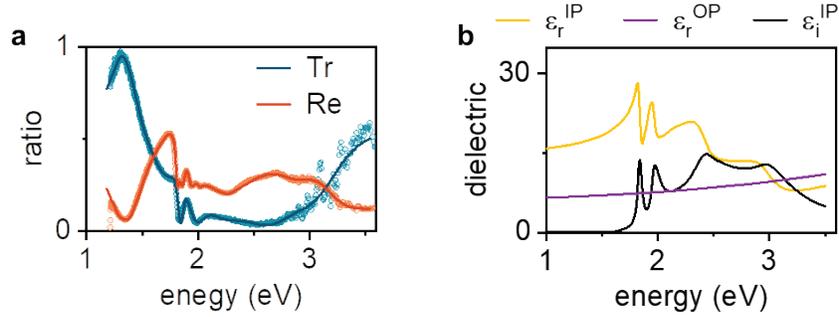

Figure S10. The dielectric function of 3$R$-MoS$_2$ is extracted from experimental reflectance and transmittance spectra. **a**. In-plane (IP) polarized transmittance (Tr, blue) and reflectance (Re, red) spectra of a 110 nm thick 3$R$-MoS$_2$ slab on Borosilicate substrate. The spectra of 3$R$-MoS$_2$ are normalized to the spectra of a 200 nm silver film on Borosilicate substrate. **b**. Extracted dielectric function of 3$R$-MoS$_2$. The IP-polarized dielectric function $\varepsilon_r^{IP}$ and $\varepsilon_i^{IP}$ are obtained from fitting Tauc-Lorentz oscillator model to the reflectance and transmission spectra, with fitting parameters sourced from Xu *et al.*[9] The out-of-plane dielectric function $\varepsilon_r^{OP}$ is referenced from Ermolaev *et al.*[10] A coupled oscillator model fit of the experimental dispersion (Figure S11, yellow and blue lines) provides a similar value of dielectric function ($\varepsilon_r^{OP}$ = 8.5) as the reference.

## 9. Polarization dependent dispersion and transport

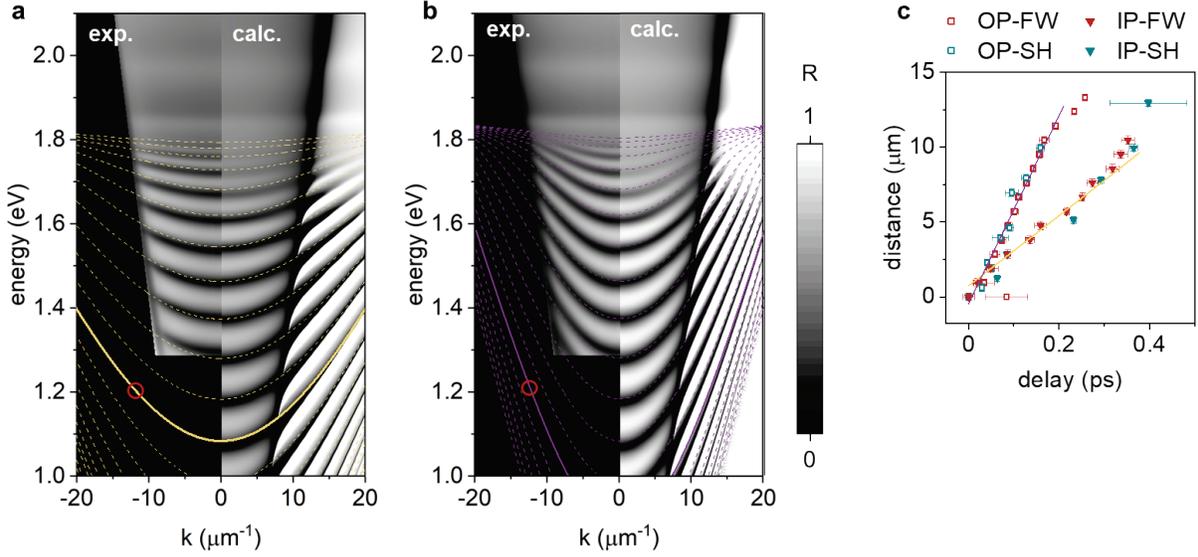

Figure S11. (a–b) Dispersion relation of 3$R$-MoS$_2$ waveguide for in-plane (IP) and out-of-plane (OP) polarization. The left panel shows experimental angle-resolved reflectance spectrum, right panel presents the transfer matrix method simulation result. The transfer matrix method that simulates the light reflectance in the SiO$_2$/MoS$_2$/immersion oil three-layer structure follows the same simulation approach in Xu $et$ $al$.[9] The yellow and purple dashed lines over the angle-resolved spectra represent cavity dispersion relation fitted by coupled oscillator models following the same approach in our previous work[1,11]. The coupled oscillator model is described by a 2N-dimension block-diagonal Hamiltonian for N waveguide modes:

$$\begin{bmatrix} U_{\text{A-ex}} & 0 & \delta_0^A & \delta_1^A & \cdots \\ 0 & U_{\text{B-ex}} & \delta_0^B & \delta_1^B & \cdots \\ \delta_0^A & \delta_0^B & H_0 & 0 & \cdots \\ \delta_1^A & \delta_1^B & 0 & H_1 & \cdots \\ \vdots & \vdots & \vdots & \vdots & \ddots \end{bmatrix} \begin{bmatrix} \chi_{\text{A-ex}} \\ \chi_{\text{B-ex}} \\ \varphi_0 \\ \varphi_1 \\ \vdots \end{bmatrix} = E_{pol} \begin{bmatrix} \chi_{\text{A-ex}} \\ \chi_{\text{B-ex}} \\ \varphi_0 \\ \varphi_1 \\ \vdots \end{bmatrix} \quad (2)$$

where $U_{\text{A-ex}}$ and $U_{\text{B-ex}}$ are the energy of the A- and B-exciton states at 1.85 eV and 2.03 eV, respectively. $H_n$ is the energy of each waveguide mode, and $\delta$ is the coupling strength between the exciton states and the waveguide modes. The fitting provides Rabi splitting of 284 meV for both exciton states. The yellow and purple solid lines highlight the main branch excited by our FW pump (see main text for details). (c) Propagation of waveguided FW (red) and SH (green) in IP-polarization and OP-polarization. Error bars are one standard deviation. The FW and SH in the same polarization share the same propagation speed. The IP-polarized waveguided light show slower propagation speed of ~8% $c$, compared to 21% $c$ for OP-polarized waveguided light, due to the larger refractive index for IP polarization, as show in figure S10b.

Note that the coupled oscillator model used to fit the single-mode cavity in Figure 4c of the main text differs from the multimode cavity in a few aspects. Because the light fields of the high momentum waveguide modes extend out from the thin 3$R$-MoS$_2$ flake, the limited spatial overlap between the light fields and the exciton states reduces their coupling strength ($\delta^A = \delta^B = 145$ meV). However, the polariton states hybridize with the surface waves as well, exhibiting anti-crossings near the TIR line. Although the surface wave hybridization is negligible in multimode cavities, it is enhanced in spatially confined singe mode waveguides ($\delta_{SW} = 60$ meV), which was observed in other exciton-polariton systems.[12] Therefore, the coupled oscillator model of the single mode cavity needs to include this surface wave hybridization in the Hamiltonian:[13]

$$\begin{bmatrix} U_{A\text{-ex}} & 0 & \delta^A & 0 \\ 0 & U_{B\text{-ex}} & \delta^B & 0 \\ \delta^A & \delta^B & H & \delta_{SW} \\ 0 & 0 & \delta_{SW} & H_{SW} \end{bmatrix} \begin{bmatrix} \chi_{A\text{-ex}} \\ \chi_{B\text{-ex}} \\ \varphi \\ \varphi_{SW} \end{bmatrix} = E_{pol} \begin{bmatrix} \chi_{A\text{-ex}} \\ \chi_{B\text{-ex}} \\ \varphi \\ \varphi_{SW} \end{bmatrix} \quad (3)$$

where $H_{SW}$ is the energy of the surface wave at the $3R$-MoS$_2$/SiO$_2$ interface. The fitting result is shown in figure 4c in the main text.

## 10. Ballistic transport and fitting details of FW propagation in thick 3*R*-MoS₂ waveguides

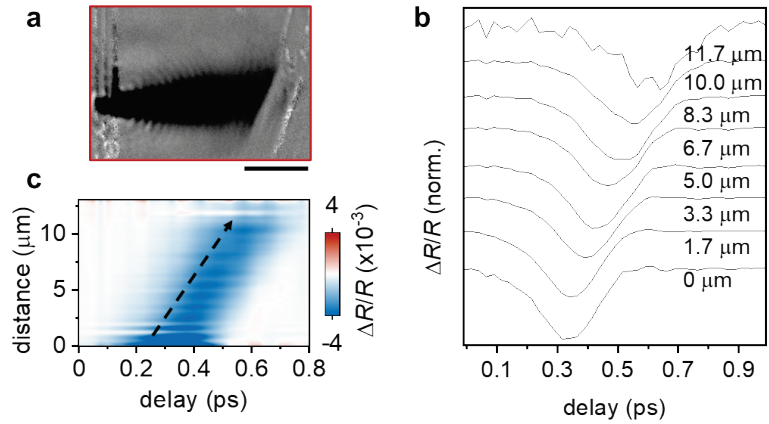

Figure S12. Ballistic transport of FW and analysis of the propagation data. (a) snapshot of the propagation of waveguided FW transport at a time delay time of 400 fs. Scale bar is 5 µm. (b) The temporal evolution of the waveguided FW stroboSCAT signal detected at distances of 0, 1.7, 3.3, 5.0, 6.7, 8.3, 10.0, 11.7 µm from the exciting edge. The FW wavepacket shows ballistic propagation with linearly increasing arrival time proportional to distance from the excitation edge. (c) Contour plot of the same ballistic transport of waveguided FW in panel b.

## 11. Anisotropic SH transport in 3R-MoS$_2$ and frequency domain characterization

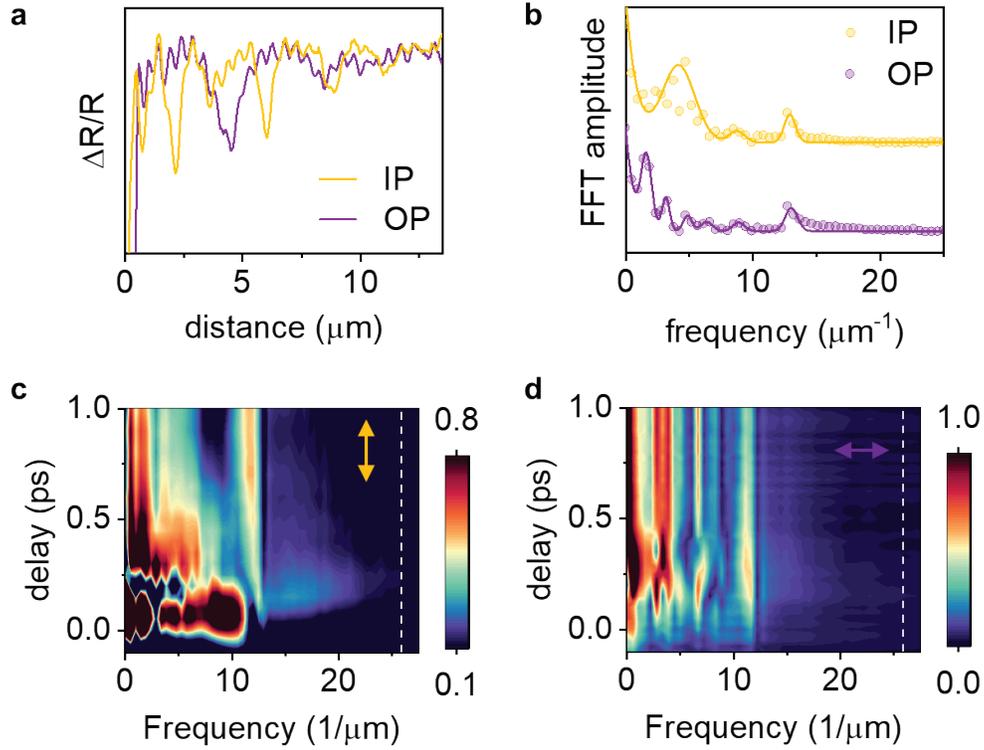

Figure S13. (a) Line-profile of SH intensity at 2 picosecond delay in IP and OP polarization for the 1.25 μm flake. (b) FFT of the spatial profile of waveguided SH in IP and OP polarization. The IP-polarized transport accommodates less modes because of the shallower dispersion (see Figure S11). (c–d). Time-resolved spatial frequency evolution of SH propagation in IP and OP polarization, respectively. The spatial frequency spectrum undergoes significant changes after ~400 fs delay. These changes occur because below-TIR modes are lossy, rapidly exiting the waveguide. At long time delays, only low-loss modes contribute to the FFT spectrum.

## 12. Maintained FW intensity throughout propagation in multimode waveguides

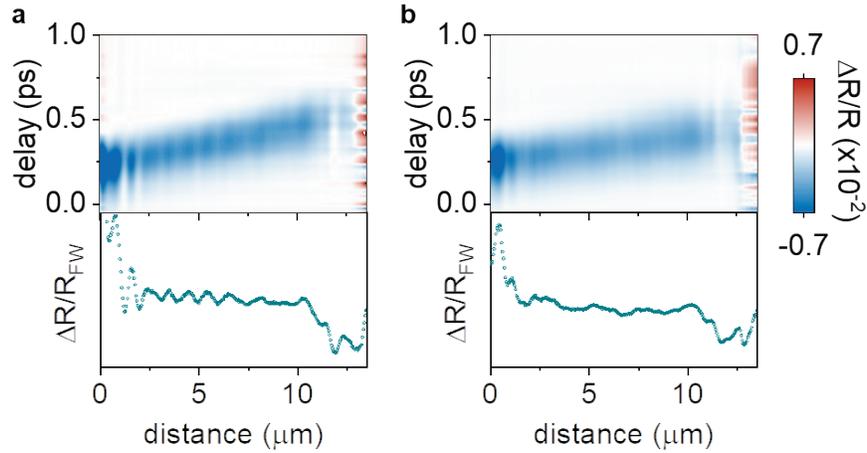

Figure S14. 2D contour plots of spatiotemporally resolved FW propagation for IP- and OP-polarizations for the 1.25 µm flake are presented in panels a and b, respectively. The rapid decay observed within the initial 2 µm of propagation results from losses from low-angle (below TIR) modes excited by edge excitation, corresponding to the pre-400 fs FFT spectra in Figure S13. The noticeable dip at 13 µm corresponds to the exit edge of the waveguide. Throughout the main part of the propagation, the intensity of the FW remains constant, indicating low losses of waveguide modes. Probe energy is at 1.66 eV. Pump fluence are 5.34 and 3.36 mJ/cm$^2$ for IP- and OP-polarization, respectively.

## 13. Numerical simulation of SH transport in lossy waveguides

The numerical simulation of the waveguide SHG transport in 3$R$-MoS$_2$ follows the theoretical work of Neuschafer et al.[14] This model focuses on a simplified scenario where the unidirectional propagating FW and SH interact in a single mode waveguide under phase-matching conditions.

The electric fields of the propagation mode at frequence ω in the z direction is:

$$E_\omega(z, t) = \frac{1}{2} U_\omega(x) E_\omega \exp[i(\omega t - vz + \phi_\omega)] \tag{3}$$

where $v$ denotes the propagation speed, and $\phi_\omega$ is a constant phase. $E_\omega$ is the constant transverse field in the one-dimensional model. $U_\omega(z)$ is the varying amplitude of the modes. For SHG under phase-matching conditions, the amplitudes of the FW and SH field, given a sufficiently long coherence length, can be calculated by using a slow-varying approximation. Thus, the amplitudes transport of FW and SH can be written as:

$$\frac{\partial}{\partial z}\begin{bmatrix} U_{FW}(z) \\ U_{SH}(z) \end{bmatrix} = \frac{1}{2}\begin{bmatrix} -\tau_{FW} & -\beta U_{FW}(z) \\ \beta U_{FW}(z) & -\tau_{SH} \end{bmatrix}\begin{bmatrix} U_{FW}(z) \\ U_{SH}(z) \end{bmatrix} \tag{4}$$

Where $\tau_{FW}$, $\tau_{SH}$ are the damping coefficients for FW and SH modes and $\beta$ is the numerical SH conversion coefficients, which is determined by the second order susceptibility $\chi^{(2)}$ and the overlap of FW and SH electric fields with the waveguide:

$$\beta \propto \int dxdy\, E_{FW}^2(x,y)\, E_{SH}(x,y)\, |\chi^{(2)}|^2 \tag{5}$$

By solving the SHG conversion matrix, the amplitudes of FW and SH are expressed as the following equations:

$$U_{FW}(z) = U_{FW}(0) \exp\{-\frac{1}{2}\int dz'[\tau_{FW} + \beta U_{SH}(z')]\} \tag{6}$$

$$U_{SH}(z) = \frac{\beta}{2}\int dz'\, U_{FW}^2(z') \exp[-\frac{1}{2}\tau_{SH}(z-z')] \tag{7}$$

The results of the fits are shown in Table S1.

| sample | $\tau_{FW}$ /μm$^{-1}$ | $\tau_{SH}$ /μm$^{-1}$ | $\beta$ /μm$^{-1}$ |
|---|---|---|---|
| 1.25 μm thick waveguide | 0 | 0.7 | 0.03 |
| 154 nm thick waveguide | 0 | 0.48 | 0.2 |
| Bulk flake[15] | NA | 28 | NA |

Table S1. Fitting parameters of the SH loss numerical simulation for the 1.25 μm and 154 nm thick waveguides. The damping coefficients of the waveguided SH are compared with the light absorption coefficients in MoS$_2$ at the same energy under normal incidence,[15] which are two orders of magnitude greater than those of the waveguided SH.

These fits can be realized on both peak intensity measurements and radially-integrated measurements with identical damping coefficients and conversion coefficients (accounting for small radial expansion of the FW and SH waves), as shown in Figure S13.

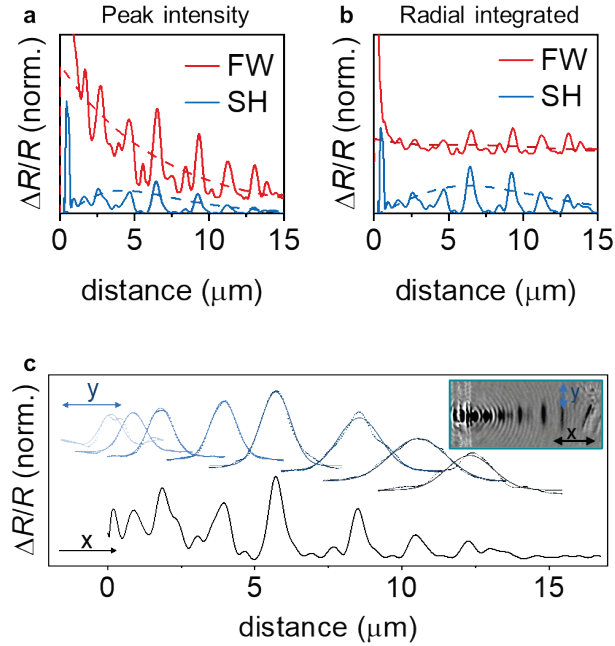

Figure S15. Numerical simulation of waveguided SHG under BPM. **a**–**b**. The intensity of waveguided FW and SH propagation are presented in peak intensity profiles (a) and radial integrated intensity profile (b), respectively. The dashed lines represent the result of numerical simulations. **c**. The expansion of waveguided SH in the transverse direction. Inset is a snapshot of a stroboSCAT image. The black curve presents the peak intensity of waveguided SH along the x direction. The blue curves present the width of the discrete stripes in the y direction. In the y direction, the width of waveguided SH expands due to the finite deviation in the scattering angle of the waveguided modes launched by edge-scattering.

## 14. Analysis of FW and SH transport in single mode 3R-MoS$_2$ waveguides

This section discusses the analysis details for measurements of SH generation in the 154 nm thick waveguide presented in the main text. Figure S16a shows the calculated dispersion, and Figure S16b shows the propagation of the FW wavepacket for different pump energies (color-coded to Figure S16a) extracted from stroboSCAT measurements. Figure S16c compares the theoretical group velocity (solid line) with the experimentally-extracted velocity. The results show excellent agreement between the theoretically-predicted group velocity from the mode dispersion and the experimentally-extracted velocity of the FW. The results also confirm that only IP-polarized modes in the TE$_0$ branch are efficiently launched by edge-scattering in this waveguide.

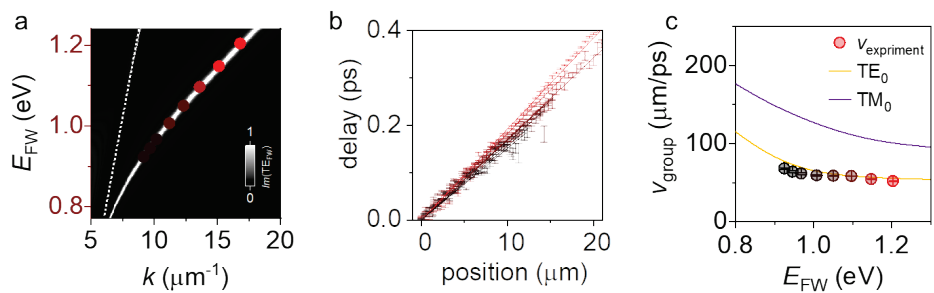

Figure S16. Dispersion relation of a single mode waveguide. (a) Dispersion relation of TE$_0$ mode in the single mode waveguide calculated by TMM simulation. (b) Linear fits of the ballistic transport of waveguided modes at pump energies of 1.20, 1.15, 1.10, 1.05, 1.01, 0.97, 0.95, and 0.92 eV. The lines of different modes are color-coded to the circles in panel a. (c) Group velocity-energy dispersion of the single mode waveguide. The solid lines in yellow and purple represent the calculated velocity from the gradient of the dispersion for IP (primarily TE$_0$ mode) and OP modes (primarily TM$_0$), whereas the red circles represent the experimental group velocities extracted from panel b.

Figure S17 compares the FW spatiotemporal evolution in the phase-mismatched case at $E_{SH}$ = 2.4 eV (Figure S17a) and phase-matched transport at $E_{SH}$ = 1.9 eV (Figure S17b) in the single-mode waveguide. When phase-mismatched, the FW maintains its intensity throughout the waveguide over ~ 20 um. When phase-matched, the FW exhibits rapid decreasing in amplitude as it propagates, suggesting rapid pump depletion. We attribute this pump depletion to efficient nonlinear conversion under phase-matched conditions.

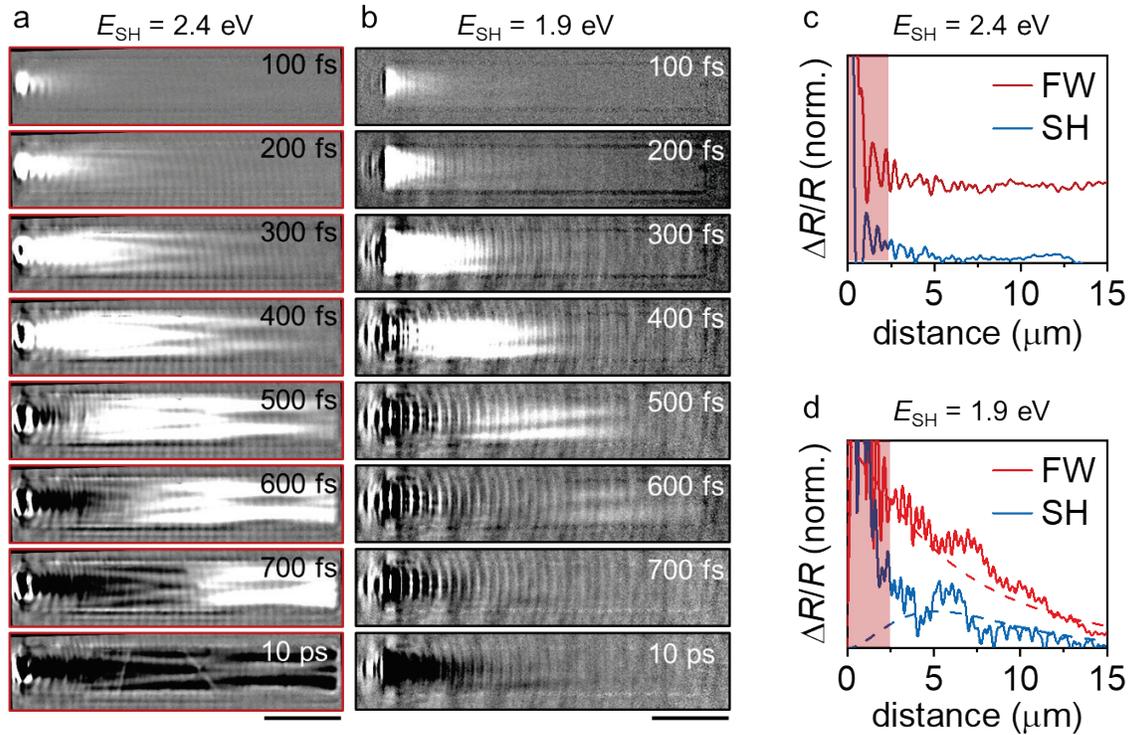

Figure S17. (a–b) Spatiotemporal imaging of waveguide FW and SH propagation in a 154 nm thick 3R-MoS$_2$ slab. SH energies are at 2.4 eV and 1.9 eV in panels a and b, respectively. The short-lived bright signal corresponds to the FW. Probe energy is at 1.96 eV. Scale bars are 5 μm. (c–d) Peak intensities of FW and SH in solid lines extracted from the measurements in panels a and b. The dashed lines represent numerical simulation results. The initial 2 μm shaded in red emphasizes the initial intense signal occurs before below-TIR (lossy) modes leak out of the waveguide.

## 15. Maintained FW intensity at $E_{SH}$= 1.9 eV in multimode waveguides

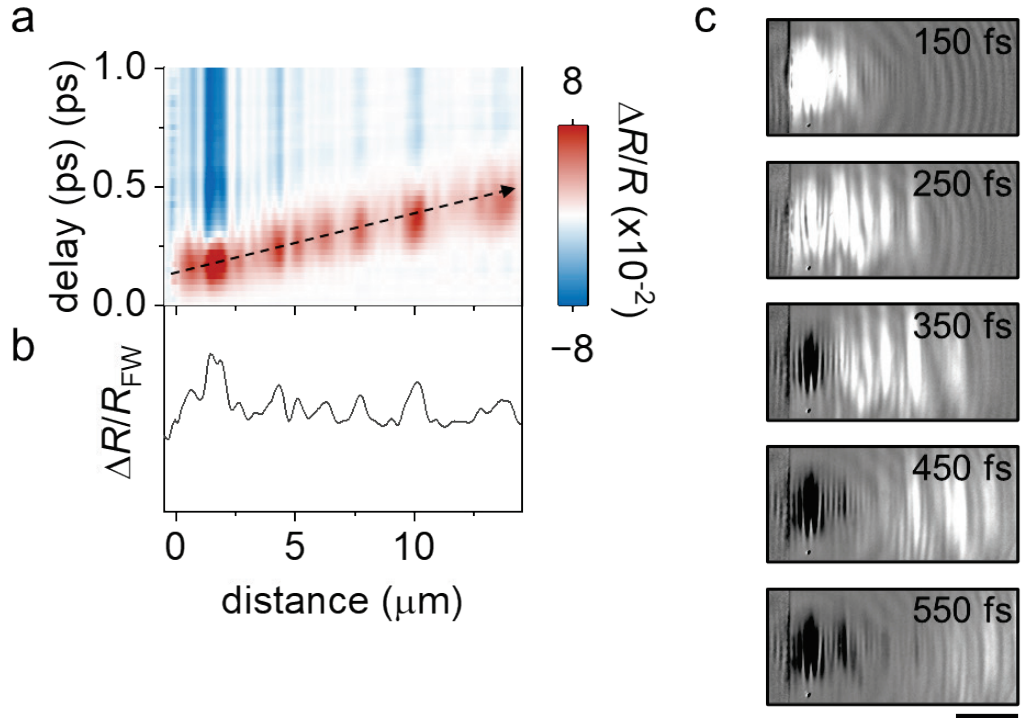

Figure S18. (a) Spatiotemporal imaging of IP-polarized FW propagation at $E_{FW}$= 0.95 eV ($E_{SH}$= 1.9 eV) in the 1.25 μm 3$R$-MoS$_2$ slab. The bright signal tracks the FW wavepacket propagation. Probe energy is at 1.66 eV. Incident pump fluence is 9.74 mJ/cm$^2$. (b) 2D contour plots of spatiotemporally resolved IP-polarized FW propagation in the slab. (c) Raw stroboSCAT images showing the FW signal intensity as a function of propagation distance. All panels show that the FW signal intensity remains approximately constant in the bulk 3$R$-MoS$_2$ slab through more than 13 μm. This observation is in stark contrast to the FW signal at the same energy in the thin slab, which vanishes within 10 μm in Figure S17b, indicating significantly stronger SH conversion in the thin 3$R$-MoS$_2$ slab thanks to polariton-assisted modal phase-matching that is absent in the thick slabs.

## 16. Frequency-domain analysis of phase-matching in thin waveguides

In this section, we discuss an approach for evaluating the phase-matching conditions through a frequency-domain analysis of fringes observed in stroboSCAT data. The SH intensity is given by:[16]

$$I_{SH}(L) \propto |\chi^{(2)}|^2 I_{FW}^2 \, \text{sinc}^2(\frac{\Delta k L}{2}) L^2 \qquad (8)$$

where $L$ is propagation distance, $I_{SH}$ and $I_{FW}$ is the intensity of SH and FW. The wavevector mismatch $\Delta k$ is defined as:

$$\Delta k = 2k_{FW} - k_{SH} \qquad (9)$$

where $k_{SH}$ and $k_{FW}$ are the momenta of SH and FW. Under phase-matching condition, $\Delta k = 0$. The periodic fringes observed in stroboSCAT measurements of the SH intensity distribution report on the convolution of $\Delta k$ with $I_{FW}^2(k)$. The fourier transform of this intensity profile is:

$$\mathcal{F}\{I_{SH}\}(k) = \mathcal{F}[\text{sinc}^2(\frac{\Delta k L}{2})L^2] * \mathcal{F}[I_{FW}^2] = \int \frac{\sqrt{2\pi}}{\Delta k^2}[2\delta(k)-\delta(k-\Delta k)-\delta(k+\Delta k)] \, \mathcal{F}\{I_{FW}^2\}(k'-k) \, dk \qquad (10)$$

where $\delta$ is the delta function, and $k$ and $k'$ are spatial frequency variables. Here we perform this analysis only for single-mode waveguide, because in multimode waveguides, $I_{FW}^2(k)$ possesses a complex spectrum due to the large number of interfering modes. The SH-FFT spectrum displays discreet bands at spatial frequency $k = 0$ and $\pm \Delta k$. Figure S19 shows phase-mismatched transport in a 146 nm thick single mode $3R$-MoS$_2$ slab waveguide, excited with the edge-scattering method as in the main text. Figure S19a, b perform the same dispersion and transport analyses as for other waveguides, allowing us to infer that the dominant mode excited by $E_{FW} = 1.203$ eV through edge excitation has a wavevector of $k_{FW} = 17.3$ μm$^{-1}$. Figure S19c shows a contour plot of the SH spatiotemporal evolution under these conditions, showing clear intensity fringes. The corresponding FFT shown in Figure S19d shows a single peak at $k = 12.1$ μm$^{-1}$. By combining these measurements, we extract a phase-matching condition determined by $k_{SH} = 2k_{FW} - \Delta k = 22.4$ μm$^{-1}$, corresponding to a phase matching angle is $\theta_{exp}^{mismatch} = 24.1°$, matching closely to that calculated using other methods in the main text.

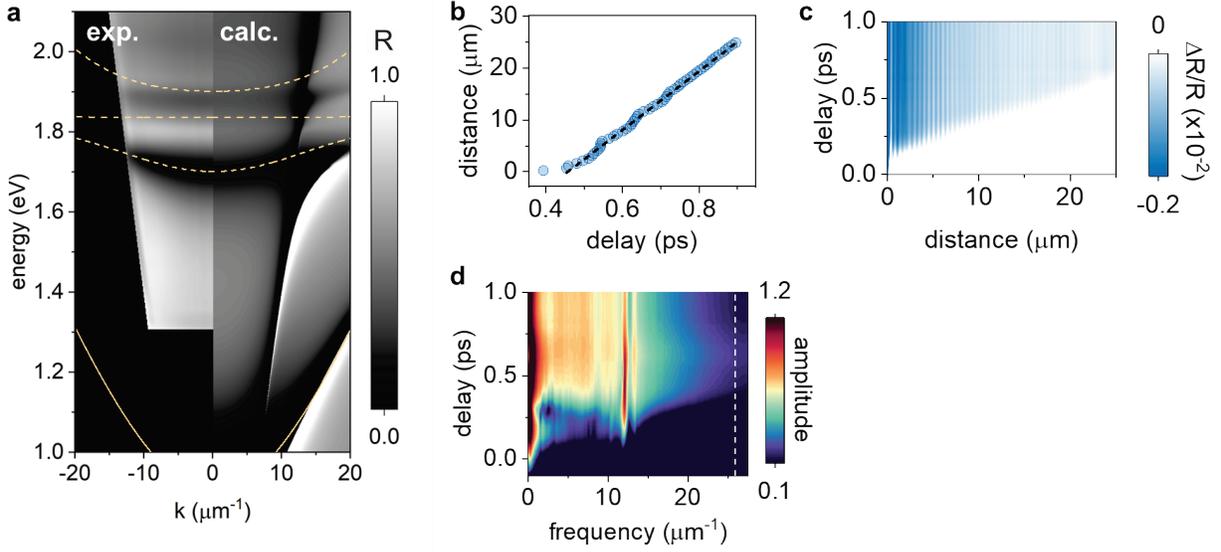

Figure S19. (a) Dispersion for a 146 nm thick 3*R*-MoS$_2$ waveguide in TE-polarization. The left and right panel shows experimental angle-resolved reflectance spectrum and the transfer matrix method simulation result, respectively. (b) coherent propagation of FW at 1.203 eV. Linear fitting shows a propagation speed of 56.7 µm/ps. (c) Spatiotemporal evolution of the SH intensity profile. (d) Time-resolved FFT spectra of the SH evolution.

To substantiate the analysis carried out above, we executed a numerical simulation of the absorption-modulated phase-mismatched transport within the single mode SHG based on eq.8:

$$I_{FW} = [\sum G(k)e^{i(k_{FW}L-\omega t)}]^2, \quad G(k) = \frac{1}{\sigma\sqrt{2\pi}}e^{-\frac{(k_{FW}-k_0)^2}{2\sigma^2}} \quad (11)$$

where $I_{FW}$ is the intensity of the electric field of FW, $k_{FW}$ and $\omega$ is the momentum and frequency of FW. The initial condition of the FW is a Gaussian wave packet defined by the pump energy width and cavity dispersion as shown in figure S20b. $G(k)$ represents the gaussian distribution of FW with a finite momentum width $\sigma = 0.1$ µm$^{-1}$. The central momentum of the FW is $k_0 = 17.7$ µm$^{-1}$. The SH intensity is then calculated with the absorption-modulated wavevector mismatched model:

$$I_{SH}^{Beer\text{-}Lambert}(L) = K\, e^{-A*L} I_{FW}^2 \operatorname{sinc}^2\left(\frac{\Delta k L}{2}\right) L^2 \quad (12)$$

where $\Delta k$ is the wavevector mismatch calculated by $\Delta k = 2k_{FW} - k_{SH}$, $K$ is weighting constant, and $A$ is the SH absorptivity. The frequency spectra are calculated by applying a FFT to the simulated spatial intensity profile of SH $I_{SH}^{Beer\text{-}Lambert}(L)$. The frequency spectrum exhibits a sharp peak corresponding to the wavevector mismatch $\Delta k$. In addition, absorption contributes to increasing spectral amplitude at lower frequencies. To account for the limited numerical aperture of our optical system, we set a high frequency cut-off to the FFT because spatial features with structures finer than λ/2NA are not resolved in our data. Therefore, all waveguided FW field components, where FFT $\mathcal{F}\{I_{FW}^2\}(k) > 40$ μm$^{-1}$, do not appear in the frequency spectrum.

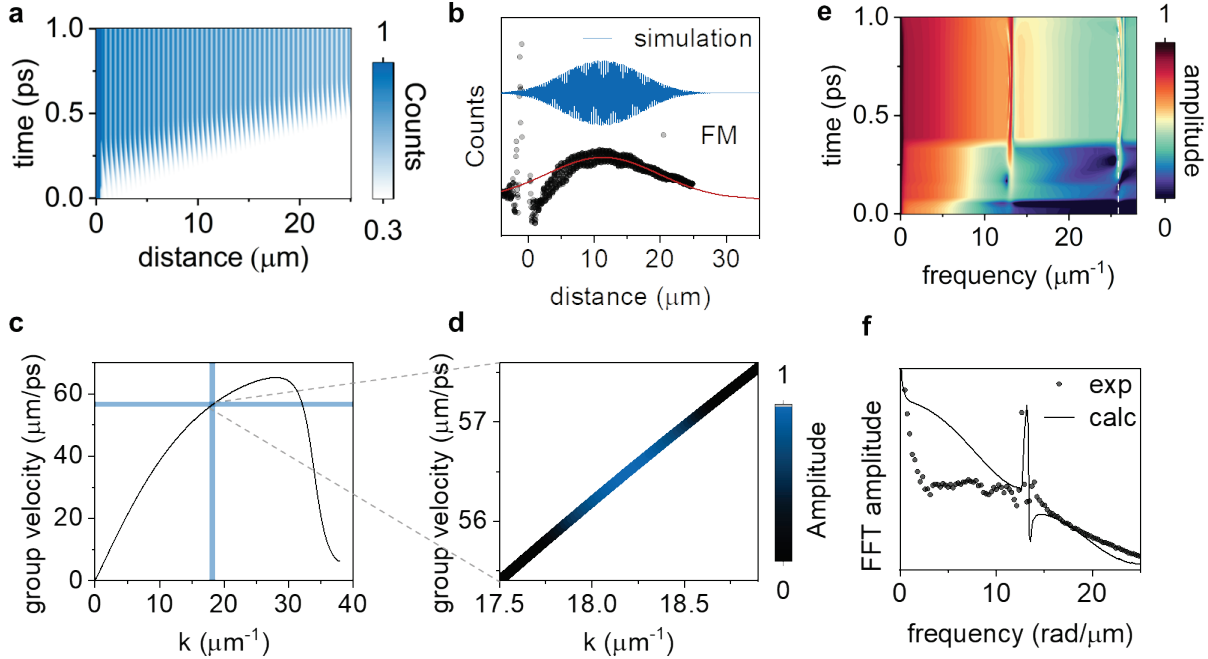

Figure S20. (a) Simulated propagation of SH under phase-mismatched conditions. (b) The spatial width of excitation in simulation and experiment. The pulsed FW propagates in the waveguide as a single wave-packet with a spatial width of 13 μm. (c) Group velocity dispersion of FW. The blue band highlights the momentum spread of the waveguided FW and the corresponding group velocity range. (d) Enlarged group velocity dispersion of the waveguided FW. The contribution of FW in different momenta distribute as a gaussian function. (e) Simulated time-resolved FFT spectra of SH in phase-mismatched transport. (e) Spatial FFT of the SH at delay time of 1 ps. The experimental and simulated FFT of phase-mismatching SH in single mode is plotted as black line.

The result aligns exceedingly well with the experimental data in both spatial and frequency domain (Figure S20a, S20e and S20f). To validate the prediction of wavevector mismatch through frequency analysis, a control simulation is conducted with varying phase-matching condition $k_{SH}$ and input central FW momentum $k_{FW}$ (Figure S21). As expected, we observe a linear relation between the peak frequency observed in the FFT contour with $k_{SH}$ and $k_{FW}$, confirming that $\Delta k$ can be directly extracted from imaging the SH intensity evolution using stroboSCAT.

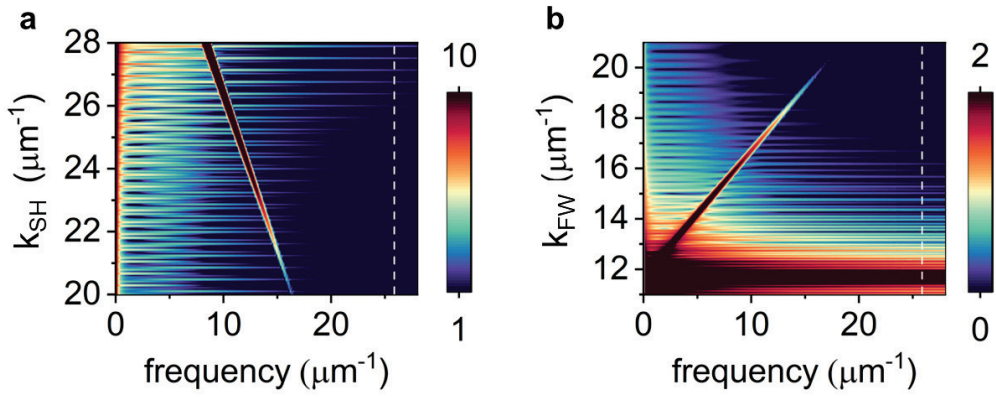

Figure S21. Numerical simulation result of waveguide mismatch with varying phase matching condition (a) and FW momenta (b).